\documentclass[a4paper,12pt]{article}
\pdfoutput=1
\usepackage{epsfig,graphicx,xcolor,amsbsy,amssymb,latexsym,amsfonts,amsmath,,tcolorbox,setspace}
\usepackage{pstricks}
\usepackage{color}
\usepackage{soul}
\usepackage[numbers,square,comma, compress]{natbib}
\usepackage{placeins}
\usepackage{wrapfig,framed,caption}
\usepackage[makeroom]{cancel}

\usepackage{wasysym}

\usepackage{multirow}

\definecolor{shadecolor}{gray}{0.925}

\usepackage{eurosym}
\usepackage[a4paper]{geometry}
\usepackage{hyperref}
\usepackage{graphicx}
\usepackage{tikz}
\usetikzlibrary{decorations.pathreplacing,decorations.markings,snakes}
\usetikzlibrary{decorations.pathmorphing}
\usetikzlibrary{patterns}
\usetikzlibrary{calc}
\usepackage{xcolor}
\usepackage{amsmath,amssymb,amsfonts,pstricks,setspace}
\usepackage[enableskew]{youngtab}

\numberwithin{equation}{section}

\newcommand{\bea}{\begin{eqnarray}\displaystyle}
\newcommand{\eea}{\end{eqnarray}}

\newcommand{\Ic}{I_{\text{c}}}

\newcommand{\Icn}[1]{I_{\text{c},#1}}
\newcommand{\ms}{S}
\newcommand{\mi}{I}
\newcommand{\mv}{V}
\newcommand{\mr}{R}

\newcommand{\rin}{\gamma}
\newcommand{\rvac}{\rho}
\newcommand{\riv}{\zeta}
\newcommand{\rhe}{\epsilon}

\newcommand{\sig}{\sigma}

\newcommand{\lam}{\lambda}

\newcommand{\herd}{h}
\newcommand{\hit}{\herd^{\text{HIT}}}

\newcommand{\Icnew}{I^{\text{new}}}

\newcommand{\cov}{SARS-Cov-2}

\newcommand{\Icml}{I^{(\text{VT})}_{\text{c}}}
\newcommand{\Icms}{I^{(\text{V})}_{\text{c}}}

\newcommand{\ep}{HP}
\newcommand{\vep}{V-HP}
\newcommand{\vtep}{VT-HP}
\newcommand{\pvt}{p_{\text{VT}}}
\newcommand{\pv}{p_{\text{V}}}

\newtcolorbox{summary}[2][]{colbacktitle=gray!40!white, colback=gray!10!white,coltitle=black, title={#2},fonttitle=\bfseries,#1}

\title{
\begin{flushright}{\vspace{-2.5cm}\small LYCEN 2021-XX\\}\end{flushright}
\vspace{2.3cm}
{\bf XXX}\\[40pt]}

\author{\large Giacomo~Cacciapaglia$^{1,2}$\footnote{\tt g.cacciapaglia@ipnl.in2p3.fr},\;
Stefan~Hohenegger$^{1,2}$\footnote{\tt s.hohenegger@ipnl.in2p3.fr},\; and Francesco Sannino$^{3,4,5}$\footnote{\tt sannino@cp3.sdu.dk}\; }

\title{
\begin{flushright}{\vspace{-2.5cm}\small LYCEN 2021-04\\}\end{flushright}
\vspace{2.3cm}
{\bf  {Effective Mathematical Modelling of Health Passes during a Pandemic}}\\[40pt]}

\begin{document}

\maketitle
\thispagestyle{empty}

\begin{center}
\renewcommand{\thefootnote}{\fnsymbol{footnote}}\vspace{-0.5cm}
${}^{1}$ Institut de Physique des 2 Infinis (IP2I) de Lyon, CNRS/IN2P3, UMR5822, \\ 69622 Villeurbanne, France\\[0.5cm]
\renewcommand{\thefootnote}{\fnsymbol{footnote}}\vspace{-0.5cm}
${}^{2}$ Universit\' e de Lyon, Universit\' e Claude Bernard Lyon 1, 69001 Lyon, France\\[0.5cm]
\renewcommand{\thefootnote}{\fnsymbol{footnote}}\vspace{-0.5cm} 
${}^{3}$  Scuola Superiore Meridionale, Largo S. Marcellino, 10, 80138 Napoli NA, Italy  \\[0.5cm]
\renewcommand{\thefootnote}{\fnsymbol{footnote}}\vspace{-0.5cm}
${}^{4}$ Dipartimento di Fisica, E. Pancini, Univ. di Napoli, Federico II and INFN sezione di Napoli, \\  Complesso Universitario di Monte S. Angelo Edificio 6, via Cintia, 80126 Napoli, Italy 
\renewcommand{\thefootnote}{\fnsymbol{footnote}}\vspace{-0.5cm} \\[0.5cm]
${}^{5}$ CP$^3$-Origins and D-IAS, Univ. of Southern Denmark,  Campusvej 55, DK-5230 Odense, Denmark
 \end{center}

\begin{center}
{\bf Abstract:} 
\end{center}
We study the impact on the epidemiological dynamics of a class of restrictive measures that are aimed at reducing the number of contacts of individuals who have a higher risk of being infected with a transmittable disease. Such measures are currently either implemented or at least discussed in numerous countries worldwide to ward off a potential new wave of COVID-19 across Europe. They come in the form of Health Passes (\ep), which grant full access to public life only to individuals with a certificate that proves that they have either been fully vaccinated, have recovered from a previous infection or have recently tested negative to \cov . We develop both a compartmental model as well as an epidemic Renormalisation Group approach, which is capable of describing the dynamics over a longer period of time, notably an entire epidemiological wave. Introducing different versions of \ep s in this model, we are capable of providing quantitative estimates on the effectiveness of the underlying measures as a function of the fraction of the population that is vaccinated and the vaccination rate. We apply our models to the latest COVID-19 wave in several European countries, notably Germany and Austria, which validate our theoretical findings.

\pagebreak

\tableofcontents
\newpage

\begin{summary}{Highlights}
{\bf Objective} To develop an effective mathematical model of Health Passes for infectious diseases. \\

\noindent
{\bf Method} We exploit the synergy between a compartmental model (SIIRV) with time-dependent rates and the epidemiological Renormalisation Group (eRG) approach to characterise an entire epidemiological wave. \\

\noindent
{\bf Main results} We demonstrate that different versions of Health Passes lead to an exponential reduction in the total number of infected individuals.\\

\noindent
{\bf Impact for healthcare and society}  Our work allows to quantify the healthcare benefits of imposing various degrees of Health Passes. For example, we find that introducing a Pass on vaccinations and negative-tests quenches the current pandemic wave, if social interactions of individuals without it are reduced by more than 30\%. Giving Health Passes to  vaccinated individuals only, makes it twice as effective. Overall, we observe a dramatic suppression in the total number of infected and a moderate flattening of the curve of new infections. Henceforth, Health Passes are powerful tools to allow societies under siege of a pandemic to reopen.
\end{summary}

\section{Introduction}
The epidemiological dynamics of \cov\, in many countries has been characterised by several waves. These are periods of exponential growth in the number of infected individuals, followed by (quasi-)linear growth phases. Modelling this dynamics in 2021 is involved due to a number of different factors: 
\begin{itemize}
\item[\emph{(i)}] After several vaccines have been developed in the second half of 2020 and have become available in large quantities, national vaccination campaigns have started in the beginning of 2021. As of August 2021, roughly 12\% of the global population has been fully vaccinated, with only a few countries reaching a rate of $>50\%$. This is still largely below the so-called herd-immunity threshold, which is the rate of the population that is required to be vaccinated to effectively prevent the spread of a virus. Furthermore, none of the currently available vaccines grants complete immunity, but still leaves a certain probability of becoming infected as well as transmitting \cov .
\item[\emph{(ii)}] After numerous mutations, several different variants of \cov\, have spread all over the globe. These variants differ regarding their infection rate, but also the various vaccines show slightly different efficacies for each of them. Recent theoretical \cite{cacciapaglia2021epidemiological} and numerical studies \cite{MLvariants} suggest that waves of COVID-19 are dominated by individual variants.
\item[\emph{(iii)}] Depending on economical, social and political factors, different countries across the globe have adopted a wide range of non-pharmaceutical interventions, ranging from lockdowns to various degrees of social distancing measures. Besides their geographic diversity, these measures also evolve over time.
\end{itemize}
With the number of vaccinated adult individuals reaching more than $25-30\%$ but still staying below the herd immunity threshold (in particular for the more aggressive new Delta-variant) and in an attempt to further allow social life to return to levels similar to ones before the pandemic, many countries have discussed (and in several cases also adopted) social distancing measures that are tailored according to the threat an individual poses to infect others. Concretely, such measures require individuals to present certificates, which prove that they present a low risk of being infectious, in order to participate in the public life. Such certificates attest that the person is fully vaccinated against COVID-19 (after having received the required number of doses of an approved vaccine and a certain waiting time), has recovered from a not too distant infection or has recently tested negative for \cov. In fact, various combinations of the above are present at national level. The social measure requires to present the certificate before entering public places (restaurants, bars, museums, shopping malls \emph{etc.}), social events (concerts, theatres, \emph{etc.}), means of mass transportation (trains, airplanes, \text{etc}.) or universities and schools. Since the concrete name differs from country to country, in the following we shall refer to such certificates collectively as \emph{Health Passes} (\ep s).\footnote{Examples for concrete implementations in different European countries can be found in Appendix~\ref{App:ExamplesPass}.}

The objective of our work is to develop a model that allows to analyse the impact of different versions of \ep s on the epidemiological dynamics of an entire wave of a pandemic. Epidemiological models capable of describing the spread of an infectious disease have a long history (for a recent review from a physics perspective, see \cite{ABC}). Most approaches are based on modelling the microscopic processes of spreading the disease either using stochastic or deterministic means. Examples of the former include lattice simulations or percolation models (see \emph{e.g.} \cite{Grassberger1983,Cardy_1985} as well as \cite{Essam,Pruessner} for reviews and further references), while the latter mostly comprise so-called \emph{compartmental models}, which were first introduced in \cite{Kermack:1927} (we refer the reader to \cite{HETHCOTErev,ABC} for further references). These models divide the population into several compartments: among others \emph{susceptible} individuals (who may become infected by the disease), \emph{infectious} individuals (who carry and spread the disease) and \emph{removed} individuals (who are no longer capable of infecting others). The epidemiological dynamics is captured by a set of coupled first order differential equations, which describe how individuals pass from one compartment to another, with fixed rates. Models of this type can easily be adapted and extended by adding compartments and are useful in establishing qualitative relations between microscopic aspects of the spread of the disease among individuals and more macroscopic observables, such as the total number of individuals who have become infected until the end of an epidemic wave. 

However, due to the fact that the epidemiological situation changes as a function of time (as we explained above) it is difficult to accurately describe the spread of a disease such as COVID-19 over a longer period (notably an entire wave, which lasts typically several weeks to months). Compartmental models with time-independent transition rates for individuals from one compartment to the next typically fail. Modelling time-dependent rates without additional input, however, typically largely reduces the predictive power of the model. Hence, in  \cite{DellaMorte:2020wlc,DellaMorte:2020qry} a complementary approach, called the \emph{epidemiological Renormalisation Group} (eRG), has been advocated, which is inspired by the physical concepts of (time-invariance) symmetry and fixed points (see \cite{Wilson1,Wilson2,Banks} for works in the context of statistical and particle physics, where the concept of Renormalisation Group Equations (RGE) has been developed). Concretely, the eRG takes the form of a set of flow equations (called the $\beta$-functions) that describe the evolution of an epidemiological quantity (\emph{e.g.} a smooth monotonic function of the cumulative number of infected individuals $\Ic$) as the flow between different fixed points. In this picture it was proposed in \cite{cacciapaglia2020evidence} that the above mentioned linear growth phase between two waves of COVID-19 corresponds to cases where this flow comes very close to a fixed point, but cannot quite reach it. This interpretation was further refined in \cite{cacciapaglia2021epidemiological} by arguing that these (quasi)fixed point are related to the interplay of different variants of the virus. Besides the computational simplicity (in its simplest form, the $\beta$-functions only depend on 2 parameters), the eRG approach is also capable of describing (and predicting) the epidemiological dynamics over a longer period of time \cite{cacciapaglia2020second}, notably an entire (or even several \cite{cacciapaglia2020evidence}) waves of COVID-19. Thus it is the preferred choice in order to model and predict the long-term time-structure of the epidemiological data.
Finally, in \cite{cacciapaglia2020us} it was further proposed how to generalise the eRG approach to include vaccinations.

In this work we shall study the effectiveness of the above mentioned \ep s by combining the flexibility of compartmental models in modelling microscopic details of the spread of a disease and their relation to more macroscopic quantities within an eRG approach. The latter efficiently encapsulates the symmetries and long-term aspects of epidemics. Concretely, we shall first introduce a so-called SIIRV model, which contains two types of people who may contract\footnote{Although the vaccines have fairly high efficacies (around 80-90\% for the best cases), the currently available vaccines do neither grant complete immunity against an infection with \cov\, nor, once infected, a transmission of the virus to others.} the disease (susceptible and vaccinated) and two types of infectious individuals (those who have been previously vaccinated and those who have not), along with the removed individuals. Next, we generalise these models by modelling measures which correspond to the introduction of a \ep. In doing so, we shall distinguish two different types of passes
\begin{itemize}
\item Vaccine and Test Health Passes (\vtep): individuals with a certificate of a negative test against \cov\, are granted the same level of access to public life as people who have been vaccinated 
\item Vaccine Health Passes (\vep): only individuals who posses a certificate for being completely vaccinated against \cov\, are granted unrestricted access to public life
\end{itemize}
In both cases, individuals that have previously contracted the disease are considered as fully immunised.
Currently, there are several examples of \vtep s implemented in various countries, while (to our knowledge) \vep s are currently only being discussed. 
We consider the two \ep\, models as extreme templates, and perform a comparative analysis of their effect on the long-term spread of the disease. One way to estimate the relative efficacy of the two models is to compare the reduction in the number of contacts for individuals who do not posses a valid pass that is necessary to obtain the same number of cumulative infected. Keeping all remaining parameters of the model to be the same, we find that these contacts need to be reduced by a factor of roughly two in \vtep\, models relative to \vep\, models to obtain the same effect. We have also found roughly the same order of magnitude in the reduction of contacts to be necessary when studying specific countries as examples.

After discussing the SIIRV model (with \ep s) we next connect it to an eRG approach. To this end, we demand that the SIIRV reproduces the solution of an eRG equation (which we shall see in examples, indeed correctly captures the time evolution of a wave of COVID-19), by allowing time dependent infection and removal rates. For the latter we find a similar functional dependence as in \cite{DellaMorte:2020qry} for a simpler class of compartmental models. Next, we generalise the compartmental models with the now time-dependent infection and removal rates by introducing the \ep s as before. The solutions we find are still solutions of the eRG equations, however, with (time-independent) parameters that are sensitive to the reduction of social contacts according to the \ep\,. This allows us to generalise the eRG approach to include \ep s, while at the same time being able to make predictions for a longer period of time than with the compartmental SIIRV. We study in detail the dependence on the \ep\, and find a simple exponential approximation for the reduction of the cumulative number of infected individuals as a function of the efficacy of the \ep . We also test this model for a number of European countries with different types of \ep s.

The remainder of this paper is organised as follows: in Section~\ref{Sect:CompartmentalSIIRV} we introduce the compartmental SIIRV model and discuss numerical and analytic properties. We also introduce the \vtep\, and \vep\, mentioned above and compare their relative efficacy. In Section~\ref{Sect:EPeRG}, by comparing to an eRG, we first generalise the SIIRV model of the previous section to time-dependent infection and removal rates. Introducing in addition the \ep\, in this time-dependent compartmental model, allows us to study how to adapt the eRG approach to include \ep s. In Section~\ref{Sect:Examples} we apply our models to some European countries: we study in detail Germany (which to date has not adapted any \ep) and Austria (which has implemented a \vtep\, model), and present relevant results for France, Denmark and Italy. Finally Section~\ref{Sect:Conclusions} contains our conclusions. Several analytical details about the SIIRV model as well as a toy model of a compartmental \ep-approach, further studies about the time-dependent infection and removal rates and details on implementations of \ep\, in several European countries are relegated to four appendices.

\section{Compartmental Vaccine Model and Health Passes}\label{Sect:CompartmentalSIIRV}
In this section we introduce a simple compartmental model that captures the basic epidemiological dynamics of an isolated population where a certain percentage of the population is immunised against the disease and a vaccination campaign for the remaining individuals is in progress. We furthermore generalise this model by introducing two versions of restrictions (via \ep s) that reduce the contacts of non-vaccinated individuals with the rest of the population in an attempt to slow the spread of the disease. This model will be used as a template to quantify the impact of the \ep s on the diffusion of the disease. 
\subsection{Compartmental Model Including Vaccinations}
\subsubsection{Basic Model}
Our starting point is an isolated population of size $N\gg 1$, which we re-group into 5 basic compartments, as listed below.
\begin{itemize}
\item Susceptible: these are individuals who are not currently infectious and who have not been (fully) vaccinated.\footnote{In the case of \cov , this also includes individuals who have only received a partial vaccination (\emph{e.g.} a single dose of Pfizer-Biontech or Astra-Zeneca without prior exposure to the virus). Here we are using recent results \cite{VaccDelta1}, which suggest that this does not provide sufficient protection against the Delta variant which is predominant, as of late summer 2021, in many countries.} They can become infectious if they come in contact with the disease. The number of susceptible individuals at time $t$ shall be denoted $N\,\ms(t)$.
\item Vaccinated: these are individuals who are not currently infectious and who are fully vaccinated. We shall, however, assume that these individuals can still get infected if they come in contact with the disease, albeit with a much smaller infection rate. We shall denote the number of vaccinated individuals as $N\,\mv(t)$.
\item Infectious without prior vaccination: these are individuals who are currently infectious (and can thus infect susceptible or vaccinated individuals if they come in contact with them) but who have not previously been vaccinated. We denote the total number of individuals in this compartment at time $t$ as $N\,\mi_1(t)$.
\item Infectious with prior vaccination: these are individuals who are currently infectious and who have previously been fully vaccinated. We shall allow for the possibility that these individuals have a reduced rate to infect other individuals than members of $\mi_1$. We shall denote the total number of individuals in this compartment as $N\,\mi_2(t)$.
\item Removed: these are individuals who can neither become infected nor can infect other individuals. This is either due to having recovered from a previous infection or by some other removal mechanism (such as quarantine or death). The number of removed individuals at time $t$ is denoted as $N\,\mr(t)$.
\end{itemize}
Individuals can pass from one of these compartments to another through various mechanisms, which we model through a number of rates. The processes are schematically shown in the following diagram:
\begin{center}
\parbox{9.8cm}{\begin{tikzpicture}
\begin{scope}[xshift=-2.5cm]
\draw[thick] (0,0) rectangle (1,1);
\node at (0.5,0.5) {$N\,\ms$};
\end{scope}
\begin{scope}[xshift=-2.5cm,yshift=-3cm]
\draw[thick] (0,0) rectangle (1,1);
\node at (0.5,0.5) {$N\,\mv$};
\end{scope}
\begin{scope}[xshift=3cm]
\draw[thick] (0,0) rectangle (1,1);
\node at (0.5,0.5) {$N\,\mi_1$};
\end{scope}
\begin{scope}[xshift=3cm,yshift=-3cm]
\draw[thick] (0,0) rectangle (1,1);
\node at (0.5,0.5) {$N\,\mi_2$};
\end{scope}

\begin{scope}[xshift=6cm,yshift=-1.5cm]
\draw[thick] (0,0) rectangle (1,1);
\node at (0.5,0.5) {$N\,\mr$};
\end{scope}
\draw[ultra thick,->] (-1.3,0.5) -- (2.8,0.5);
\node at (0.75,0.9) {$N\,(\rin_1\,\mi_1+\rin_2\,\mi_2)\,\ms$};
\draw[ultra thick,->] (-1.3,-2.5) -- (2.8,-2.5);
\node at (0.75,-2.9) {$N\,\riv\,(\rin_1\,\mi_1+\rin_2\,\mi_2)\,\mv$};
\draw[ultra thick,->] (4.2,0.5) -- (6.5,0.5) -- (6.5,-0.4);
\node at (5.2,0.9) {$N\,\rhe\,\mi_1$};
\draw[ultra thick,->] (4.2,-2.5) -- (6.5,-2.5) -- (6.5,-1.6);
\node at (5.2,-2.9) {$N\,\rhe\,\mi_2$};
\draw[ultra thick,->] (-2,-0.25) -- (-2,-1.75);
\node[rotate=90] at (-2.4,-1) {$N\,\rvac\,\ms$};
\end{tikzpicture}}
\end{center}

\noindent
Mathematically, the system can be modelled by the following coupled first order differential equations in time
\begin{align}
&\frac{d\ms}{dt}(t)=-\ms(t)\,\left[\rvac+\rin_1\,\mi_1(t)+\rin_2\,\mi_2(t)\right]\,,&&\frac{d\mi_1}{dt}(t)=\ms(t)\,\left[\rin_1\,\mi_1(t)+\rin_2\,\mi_2(t)\right]-\rhe\,\mi_1(t)\,,\nonumber\\
&\frac{d\mv}{dt}(t)=\rvac\,\ms(t)-\mv(t)\,\riv\,\left[\rin_1\,\mi_1(t)+\rin_2\,\mi_2(t)\right]\,,&&\frac{d\mi_2}{dt}(t)=\mv(t)\,\riv\,\left[\rin_1\,\mi_1(t)+\rin_2\,\mi_2(t)\right]-\rhe\,\mi_2(t)\,,\nonumber\\
&\frac{d\mr}{dt}(t)=\rhe\,\left[\mi_1(t)+\mi_2(t)\right]\,,\label{DiffSIIRV}
\end{align}
which need to be supplemented by the initial conditions
\begin{align}
&\ms(t=0)=\ms_0\,,&&\mi_1(t=0)=\mi_{1,0}\,,&&\mi_2(t=0)=\mi_{2,0}\,,&&\mv(t=0)=\mv_0\,,&&\mr(t=0)=0\,.\label{InitCondSIIRV}
\end{align}
Here we assume the outbreak of the disease at $t=0$ and we normalise the initial conditions to satisfy $\ms_0+\mi_{1,0}+\mi_{2,0}+\mv_0=1$.\footnote{Since the equations (\ref{DiffSIIRV}) imply $\frac{d}{dt}(\ms+\mv+\mi_1+\mi_2+\mr)=0$, this also means $(\ms+\mv+\mi_1+\mi_2+\mr)(t)=1$ for any $t\geq 0$.} In (\ref{DiffSIIRV}) $\rin_1$ and $\rin_2 \in\mathbb{R}_+$ are the rates at which infectious individuals with or without prior vaccination infect susceptible individuals. These two rates are not considered a priori the same (but for most examples, we shall use for simplicity $\rin_1=\rin_2$). Furthermore, $\rhe$ denotes the recovery rate, which is assumed to be independent of whether individuals have been previously vaccinated or not. We shall also use the quantities
\begin{align}
&\sig_1=\frac{\rin_1}{\rhe}\,,&&\text{and} &&\sig_2=\frac{\rin_2}{\rhe}\,,\label{DefSigs}
\end{align}
which correspond to the reproduction number of the two infectious compartments.
The rate $\rvac$ in (\ref{DiffSIIRV}), denotes the vaccination rate, which is chosen to be constant.\footnote{As we shall see in the examples discussed in Section~\ref{Sect:Examples}, this indeed seems to lead to a reasonable approximation for a single wave of COVID-19.} Finally, the efficacy of the vaccine is encoded in the reduction factor $\riv\in[0,1)$ for the infection rate of vaccinated individuals. Estimated numbers for this parameter based on recent studies for different vaccines agains \cov\, were for example found in \cite{VaccDelta1}. For later use, we also define the cumulative number of infected individuals
\begin{align}
\Ic(t)&=N\,(\mi_{1,0}+\mi_{2,0})+N\int_0^t dt'\,\left[\ms(t')+\riv\,\mv(t')\right]\,\left[\rin_1\,\mi_1(t')+\rin_2\,\mi_2(t')\right]\,,
\end{align}
as a function of time. We shall discuss numerical solutions of $\Ic$ later on 
and compare them with generalisations of the model (\ref{DiffSIIRV}), which restrict the contacts of unvaccinated individuals (see Section~\ref{Sect:GreenPassesConstantRates}).

\subsubsection{Herd Immunity}\label{Sect:HerdSIIRV}
\begin{figure}[htbp]
\begin{center}
\includegraphics[width=7.5cm]{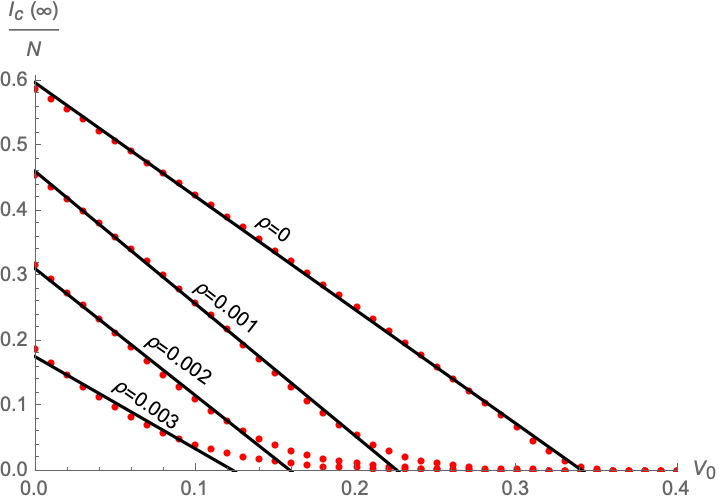}\hspace{1cm}\includegraphics[width=7.5cm]{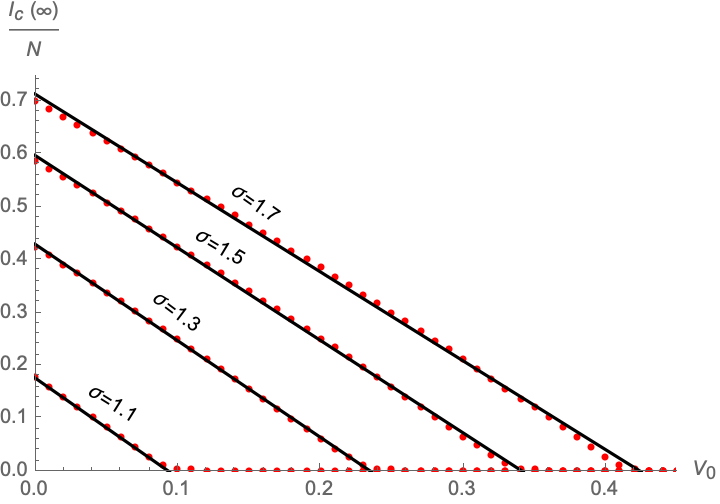}
\end{center}
\caption{Left panel: Asymptotic cumulative number of infected individuals as a function of $\mv_0$ for different values of the vaccination rate $\rho$. The plot furthermore uses $\sig_1=\sig_2=1.5$, $\rhe=0.2$ and $\riv=0$. Right panel: Asymptotic cumulative number of infected individuals as a function of $\mv_0$ for different values of $\sig_1$. The plot furthermore uses $\sig_1=\sig_2$, $\rhe=0.2$, $\rvac=0$ and $\riv=0$}
\label{Fig:HerdImmunity}
\end{figure}

An important parameter in the compartmental model (\ref{DiffSIIRV}) and (\ref{InitCondSIIRV}) is the number of vaccinated individuals at the outbreak of the disease, $V_0$. Indeed, the asymptotic behaviour of the dynamics crucially depends on it and, for fixed $\rin_{1,2}$, $\rhe$, $\rho$ and $\riv$, the infinite-time limit $\Ic(\infty)$ shows a critical behaviour with respect to this parameter: as shown in the numerical plots in Figure~\ref{Fig:HerdImmunity}, the asymptotic cumulative number of infected individuals shows a linear decrease as a function of $V_0$ up to a certain (critical) value, above which the number of infected during the entire outbreaks remains relatively small (compared to the total size of the population).

The phenomenon that the spread of the disease is severely hampered if a certain critical percentage of the population has been immunised is called \emph{herd immunity}. In the current model, we expect that this threshold depends on the various parameters. However, since the effect is most pronounced for vanishing vaccination rate ($\rho=0$), we have analysed the system for this value and obtained an analytical estimate in appendix~\ref{App:HerdImmunitySIIRV}. Concretely, in the SIIRV model (\ref{DiffSIIRV}), we define the herd immunity threshold $\hit$ as the minimal fraction of the population that needs to be vaccinated beforehand, such that during the outbreak of the disease\footnote{For the outbreak, we shall consider the limit $\mi_{1,0}\to0$ and $\mi_{2,0}\to 0$.} the number of new infections never reaches an extremum, but remains a monotonically decreasing function. The value found in (\ref{HIT}) is
\begin{align}
\hit=\frac{\sig_1-1}{\sig_1-\riv\sig_2}\,,
\end{align}
which indeed fits with the numerical plots in the right panel of Figure~\ref{Fig:HerdImmunity}. Notice that for $\riv\sig_2>1$, we formally find $\hit>1$: in this case, the efficacy of the vaccine is too low and herd immunity cannot be achieved.

\subsection{Implementing Health Passes}\label{Sect:GreenPassesConstantRates}
The factor $\riv$ in the SIIRV model (\ref{DiffSIIRV}) takes into account mainly biological effects of the various vaccines and a priori is not related to any social distancing measures particularly targeted at unvaccinated individuals. However, after the start of vaccination campaigns in the beginning of 2021, many countries have started in the summer of this year to discuss (and in some cases also implement) such measures, based on obtaining a health pass. Indeed, these measures allow access to public places and social events with a high concentration of people (such as museums, concerts, restaurants, bars, shopping malls etc.) only to individuals who can either prove a certain level of immunisation against \cov\, and/or have recently tested negative for the virus. The details and scope of these measures differ strongly among various countries, but from an epidemiological perspective, they are aimed at reducing the contacts of individuals who stand at a greater risk of being infected with the rest of the population. To implement these measures into the SIIRV model, we distinguish two conceptually different types of \ep s:
\begin{itemize}
\item Vaccine and Test Health Passes (\vtep): individuals who have tested negative for \cov\, are treated in the same way as individuals who are fully immunised, \emph{i.e.} individuals with a negative test certificate are granted the same access to public life as individuals with a certificate for a completed vaccination scheme. We furthermore assume, in this scenario, that tests are easily accessible (and free of charge) for the majority of the population. Examples for models of this type which have actually been implemented are the Austrian '3-G-Regel', the Danish 'Corona Pass' or the French 'pass sanitaire' .

From the perspective of the compartmental model (\ref{DiffSIIRV}), we can implement such restrictions through a supression factor $\pvt\in[0,1]$ that takes into account how much contacts of the unvaccinated, infected individuals $\mi_1$ with the rest of the population are reduced
\begin{align}
&\frac{d\ms}{dt}=-\ms\,\left[\rvac+\pvt\,\rin_1\,\mi_1(t)+\rin_2\,\mi_2\right]\,,&&\frac{d\mi_1}{dt}=\ms\,\left[\pvt\,\rin_1\,\mi_1+\rin_2\,\mi_2\right]-\rhe\,\mi_1\,,\nonumber\\
&\frac{d\mv}{dt}=\rvac\,\ms-\mv\,\riv\,\left[\pvt\,\rin_1\,\mi_1+\rin_2\,\mi_2\right]\,,&&\frac{d\mi_2}{dt}=\mv\,\riv\,\left[\pvt\,\rin_1\,\mi_1+\rin_2\,\mi_2\right]-\rhe\,\mi_2\,,\nonumber\\
&\frac{d\mr}{dt}=\rhe\,\left[\mi_1+\mi_2\right]\,.\label{DiffSIIRVgreenpass3G}
\end{align}
Mathematically, the introduction corresponds to a rescaling of the infection rate for unvaccinated individuals $\rin_1$. The cumulative number of infected individuals for this model becomes
\begin{align}
\Icml(t,\pvt)&=N\,(\mi_{1,0}+\mi_{2,0})+N\int_0^t dt'\,\left[\ms(t')+\riv\,\mv(t')\right]\,\left[\pvt\,\rin_1\,\mi_1(t')+\rin_2\,\mi_2(t')\right]\,.
\end{align}

\item Vaccine Health Passes (\vep): only fully vaccinated individuals are subject to no restrictions in public. This in particular implies that individuals who have (recently) tested negative against \cov, but who are not fully vaccinated, are still only allowed restricted access to public life. In practice, this model also applies to situations in which the majority of the population has no simple access to (free) tests and the only viable option to gain an \ep\, is a vaccination certificate.

Upon completion of this paper, to our knowledge, no government has fully implemented such a model in all of public life, but it is publicly discussed (\emph{e.g.} a so-called '1-G-Regel' is proposed in Austria for access to discotheques and nightclubs in fall 2021).  Recently the Biden administration is also discussing restrictions based on vaccinated individual for all levels of the American workforce and generally for a controlled reopening of the society. 
 
Similar to the model (\ref{DiffSIIRVgreenpass3G}), these conditions can also be implemented through a parameter $\pv\in[0,1]$ that measures their efficacy. However, in contrast to (\ref{DiffSIIRVgreenpass3G}), the factor also applies to contacts for the susceptible individuals (who are unvaccinated)
\begin{align}
&\frac{d\ms}{dt}=-\ms\,\left[\rvac+\pv^2\,\rin_1\,\mi_1+\pv\,\rin_2\,\mi_2\right]\,,&&\frac{d\mi_1}{dt}=\ms\,\left[\pv^2\,\rin_1\,\mi_1+\pv\,\rin_2\,\mi_2\right]-\rhe\,\mi_1\,,\nonumber\\
&\frac{d\mv}{dt}=\rvac\,\ms-\mv\,\riv\,\left[\pv\,\rin_1\,\mi_1+\rin_2\,\mi_2\right]\,,&&\frac{d\mi_2}{dt}=\mv\,\riv\,\left[\pv\,\rin_1\,\mi_1+\rin_2\,\mi_2\right]-\rhe\,\mi_2\,,\nonumber\\
&\frac{d\mr}{dt}=\rhe\,\left[\mi_1+\mi_2\right]\,.\label{DiffSIIRVgreenpass}
\end{align}
The cumulative number of infected individuals is given by
\begin{align}
\Icms(t,\pv)&=N\,(\mi_{1,0}+\mi_{2,0})+N\int_0^t dt'\,\left[\pv\,\ms(t')+\riv\,\mv(t')\right]\,\left[\pv\,\rin_1\,\mi_1(t')+\rin_2\,\mi_2(t')\right]\,.
\end{align}
\end{itemize}

\begin{figure}[h]
\begin{center}
\includegraphics[width=7.5cm]{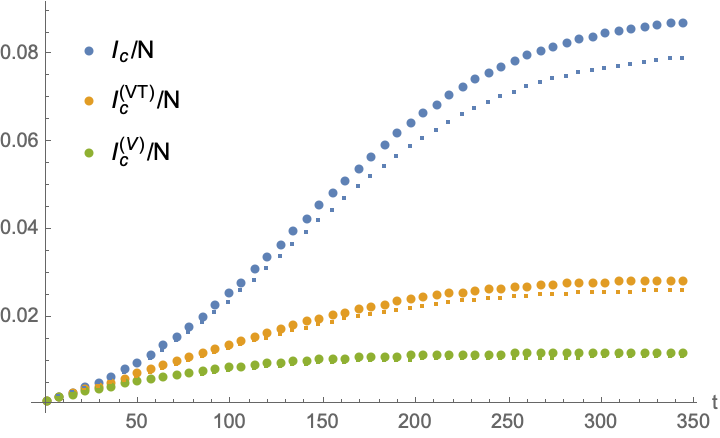}\hspace{1cm}\includegraphics[width=7.5cm]{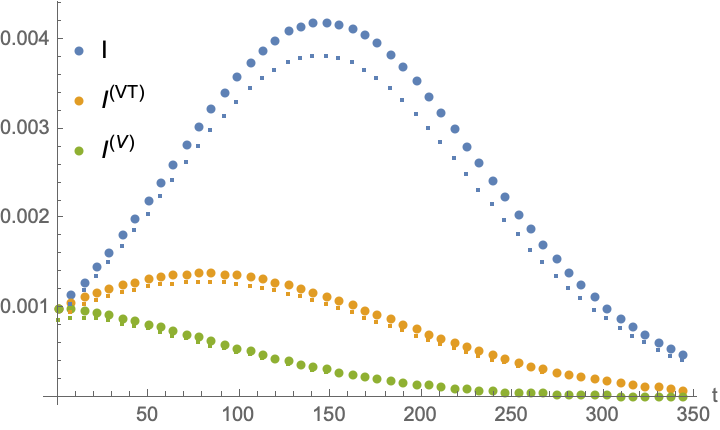}
\end{center}
\caption{Numerical solutions of the SIIRV model with different variants of a \ep: solutions of eq.~(\ref{DiffSIIRV}) are represented by blue points, those of eq.~\ref{DiffSIIRVgreenpass3G} by orange points and those of (\ref{DiffSIIRVgreenpass}) by green points. The left panel shows the cumulative number of infected individuals (large points stand for the total numbers, while small points represent only the unvaccinated individuals) and the right panel the infectious individuals as functions of time. Both plots use $\sig_1=\sig_2=1.6$, $\rhe=0.1$, $\riv=0.15$, $\rvac=0.0005$ and $\mv_0=0.3$.}
\label{Fig:ComparisonGreenPassNumbers}
\end{figure}

\begin{figure}[h]
\begin{center}
\includegraphics[width=7.5cm]{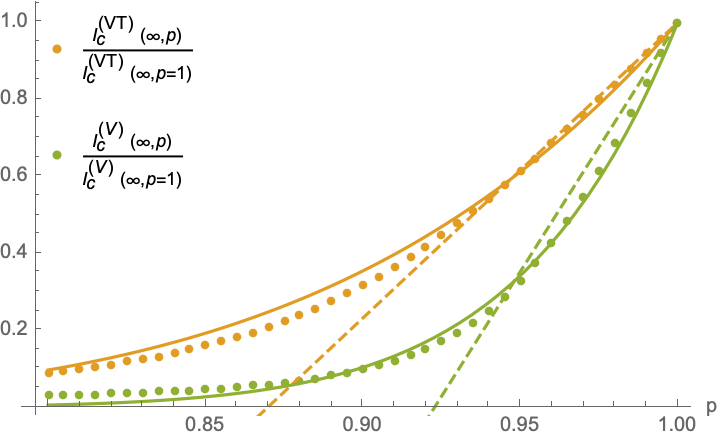}\hspace{1cm}\includegraphics[width=7.5cm]{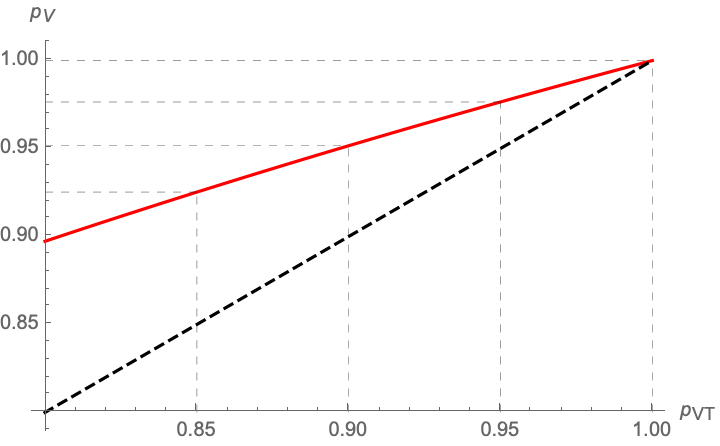}
\end{center}
\caption{Left panel: comparison of the asymptotic cumulative number of infected individuals as a function of $p$ (and normalised to $p=1$) for (\ref{DiffSIIRVgreenpass3G}) and (\ref{DiffSIIRVgreenpass}). The dots represent the numerical solutions, while the dashed lines stand for the leading (linear) approximation at $p=1$ and the solid lines for interpolations with exponential functions of the form (\ref{GreenAsyInter}). The plot uses $\sig_1=\sig_2=1.6$, $\rhe=0.1$, $\riv=0.15$, $\rvac=0.0005$ and $\mv_0=0.3$ and leads to the interpolation parameters $\theta^{(\text{VT})}=9.326$ and $\theta^{(\text{V})}=20.364$.  Right panel: with these same parameters, the red curve represents the relation between the $p$-parameters of (\ref{DiffSIIRVgreenpass3G}) and (\ref{DiffSIIRVgreenpass}) that lead to the same asymptotic cumulative number of infected individuals. The dashed black line represents for comparison the relation $\pv=\pvt$.}
\label{Fig:ComparisonGreenPassEquiv}
\end{figure}

\noindent
Note that in both scenarios, the individuals that have recovered from a recent infection are considered as fully immunised, and counted as removed individuals.
Numerical solutions of the coupled differential equations (\ref{DiffSIIRVgreenpass3G}) and (\ref{DiffSIIRVgreenpass}) are shown in Figure~\ref{Fig:ComparisonGreenPassNumbers}. The right panel of this Figure is a first hint for the potential of a \ep\, to 'flatten the curve', \emph{i.e.} to reduce the local maximum of the number of infectious individuals as a function of time, or even completely eliminate it. To further study this point, the left panel of Figure~\ref{Fig:ComparisonGreenPassEquiv} shows a comparison of the asymptotic cumulative number of infected individuals as a function of $p$ (and normalised to $p=1$) for the two models of \ep s (\ref{DiffSIIRVgreenpass3G}) and (\ref{DiffSIIRVgreenpass}). The numerical solutions can be interpolated by exponential functions of the form\footnote{In the following we shall distinguish between $\pvt$ and $\pv$ if there is a risk of confusion. In cases when we discuss them collectively (or if there is no risk of confusion), we shall simply denote $p$ the efficacy of the \ep\,.}
\begin{align}
&\Ic^{(\text{VT},\text{T})}(\infty,p)\sim \Ic^{(\text{VT},\text{T})}(\infty,p=1)\,\text{exp}\left(\theta^{(\text{VT},\text{T})}\,\frac{p-1}{p}\right)\,,&&\text{with} &&\theta^{(\text{VT},\text{T})}\in\mathbb{R}_+\,,\label{GreenAsyInter}
\end{align}
with the two models mainly differing by the constant fitting parameters $\theta^{(\text{VT})}$ and $\theta^{(V)}$.\footnote{Approximations of this type and their quality can already be analysed in a 'classical' SIR model \cite{Kermack:1927} (\emph{i.e.} without any vacciantion dynamics). For more details, including a comparison of (\ref{GreenAsyInter}) with the first and second order of a Taylor series expansion around $p=1$ can be found in Appendix~\ref{App:ComparSIR}.} Assuming all remaining parameters to be the same (notably the recovery rate $\rhe$), the same efficacy of the \ep  models (\ref{DiffSIIRVgreenpass3G}) and (\ref{DiffSIIRVgreenpass}), $\pvt$ and $\pv$ respectively, lead to different asymptotic cumulative numbers of infected individuals. We can turn this relation around by determining which values of $\pvt$ and $\pv$ (for all other parameters being held fixed), lead to the same number of infected individuals at the end of the epidemic. The red line in the right panel of Figure~\ref{Fig:ComparisonGreenPassEquiv} shows the relation between $\pv$ and $\pvt$ that needs to be satisfied in order to obtain the same asymptotic behaviour: for the parameters chosen, $\pvt$ in a \ep\, that accepts both test and vaccination certificates needs to be roughly by a factor larger than $\pv$ in a \ep\, which only allows vaccinated individuals full access to public life.

\begin{figure}[h]
\begin{center}
\includegraphics[width=7.5cm]{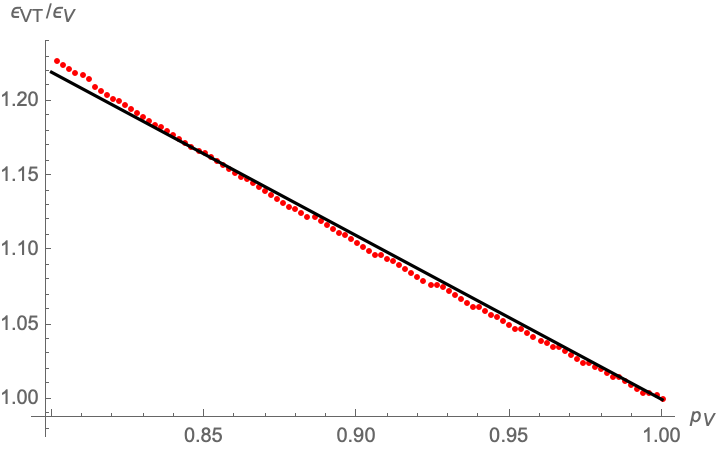}\hspace{1cm}\includegraphics[width=7.5cm]{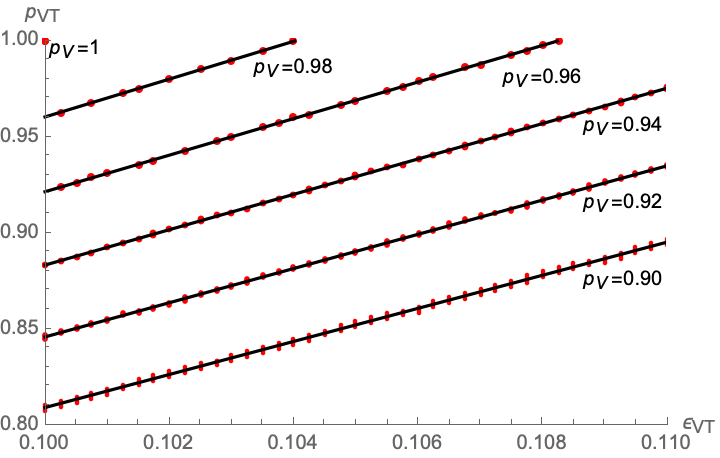}
\caption{Value of $\rhe_{\text{VT}}/\rhe_{\text{V}}$ that is necessary for given $\pv=\pvt$ to lead to the same asymptotics of (\ref{DiffSIIRVgreenpass3G}) and (\ref{DiffSIIRVgreenpass}).}
\label{Fig:ComparisonGreenPassEquivEps}
\end{center}
\end{figure}

The relation in the right panel of (\ref{Fig:ComparisonGreenPassEquiv}) can be studied using the approximation (\ref{GreenAsyInter}), which implies equivalent asymptotic numbers of infected individuals for
\begin{align}
\pv&=\frac{\pv\,\theta^{(\text{V})}}{\pvt (\theta^{(\text{V})}-\theta^{(\text{VT})})+\theta^{(\text{VT})}}\nonumber\\
&=1+\frac{\theta^{\text{(VT)}}}{\theta^{\text{V}}}\,(\pvt-1)+\frac{\theta^{(\text{VT})}(\theta^{(\text{VT})}-\theta^{(\text{V})})}{(\theta^{(\text{V})})^2}\,(\pvt-1)^2+\mathcal{O}((\pvt-1)^3)\,.
\end{align}
The linear approximation around $\pvt=1$ with the coefficient $\frac{\theta^{\text{(VT)}}}{\theta^{\text{V}}}\sim 0.458$ indeed very well agrees with the right panel of Figure~\ref{Fig:ComparisonGreenPassEquiv}.

We stress, however, that the comparison in the right panel of Figure~\ref{Fig:ComparisonGreenPassEquiv} assumes that all remaining parameters of the numerical analysis remain perfectly the same for both models. In particular, we assumed the same removal rate $\rhe$ in both cases, which (among other things) depends on the efficiency of the contact-tracing, \emph{i.e.} identifying and quarantining infected individuals and therefore also crucially depends on the number of tests that are being performed per time unit. Since the model (\ref{DiffSIIRVgreenpass}) offers less incentive for individuals to get tested (unless they present clear symptoms), it is to be expected that in this case the test rate is lower, leading ultimately to a smaller value of $\rhe$. In Figure~\ref{Fig:ComparisonGreenPassEquivEps} we have therefore plotted the relation of $\rhe_{\text{VT}}$ in (\ref{DiffSIIRVgreenpass3G}) (normalised to $\rhe_{\text{VT}}$ in (\ref{DiffSIIRVgreenpass})) that is necessary for given $\pv=\pvt$ to lead to the same asymptotics of (\ref{DiffSIIRVgreenpass3G}) and (\ref{DiffSIIRVgreenpass}).

We remark, in particular, that the right panel of Figure~\ref{Fig:ComparisonGreenPassEquivEps} suggests a linear equivalence between the parameter $\pvt$ and the removal rate $\rhe$. This can be explained by the fact that in the model (\ref{DiffSIIRVgreenpass3G}) the asymptotic cumulative number of infected individuals  depends to good approximation on the rates only in the combination $\sig_1=\frac{\rin_1}{\rhe}=\sig_2$. Furthermore, for the model (\ref{DiffSIIRVgreenpass3G}), the parameter $\pvt$ can be reabsorbed by a rescaling of $\rin_1$. This explains why a linear relation among $\pvt$ and $\rhe$ yields equivalent asymptotic results.

\section{Epidemiological Renormalisation Group}\label{Sect:EPeRG}
As it has been demonstrated in \cite{DellaMorte:2020wlc,DellaMorte:2020qry}, compartmental models with constant rates are in general not capable of describing the epidemiological evolution of an entire wave of COVID-19. Instead an epidemiological Renormalisation Group (eRG) approach was proposed which describes the spread of a disease through flow equations (the so called $\beta$-functions) and characterises a wave as the flow between fixed points  \cite{DellaMorte:2020wlc,cacciapaglia2020second,cacciapaglia2020evidence,cacciapaglia2021epidemiological}. Being based on temporal symmetries of the epidemiological dynamics, this approach was demonstrated \cite{DellaMorte:2020wlc,cacciapaglia2020second,Cacciapaglia:2021vvu} to capture accurately an entire wave. Concretely, let $\Ic(t)$ denote the cumulative number of infected individuals and $\alpha=\phi(\Ic)$ with $\phi$ a continuous, differentiable and monotonic function. The $\beta$-function for a single wave (and a single variant of a disease) can be written as
\begin{align}
-\beta_{\alpha}(t)=\frac{d\alpha}{dt}=\frac{d\phi}{d\Ic}\,\frac{d\Ic}{dt}=\lambda_0\,\alpha\,\left(1-\frac{\alpha}{A_0}\right)^{2d}\,,\label{BetaFunction}
\end{align}
with $(A_0,\lambda_0,d)$ constants. Specifically, $\lambda_0$ is related to the infection rate of the disease, while $A_0$ is the asymptotic number of individuals who get infected during the wave. For simplicity, we shall consider $d=\frac{1}{2}$ and $\alpha=\phi(\Ic)=\Ic$ in the following. In this case, the solution of the flow equation (\ref{BetaFunction}) is a logistic function
\begin{align}
\Ic(t)=\frac{A_0}{1+e^{-\lambda_0(t-t_0)}}\,,\label{LogisticFunction}
\end{align}
where $t_0\in\mathbb{R}$ is an integration constant that is related to the starting point of the onset of the wave. As demonstrated in \cite{DellaMorte:2020wlc,DellaMorte:2020qry,Cacciapaglia:2020mjf,cacciapaglia2020second,cacciapaglia2021epidemiological} (and as we shall see in the examples below), for suitable parameters $(A_0,\lambda_0,t_0)$, the function (\ref{LogisticFunction}) indeed describes accurately the time evolution of infected individuals during a single wave of COVID-19 even for populations that differ greatly geographically as well as socio-culturally and under very different circumstances regarding non-pharmaceutical interventions, vaccines and variants of \cov\,. In the following we shall explore ways to obtain solutions of this type (which describe the time evolution of an entire wave) from the compartmental model (\ref{DiffSIIRV}), which can be later combined with \ep s along the lines of (\ref{DiffSIIRVgreenpass3G}) and (\ref{DiffSIIRVgreenpass}).
\subsection{Time Dependent Rates and Relation to SIIRV}\label{Sect:TimeDepSIIRV}

\begin{figure}[htbp]
\begin{center}
\includegraphics[width=7.5cm]{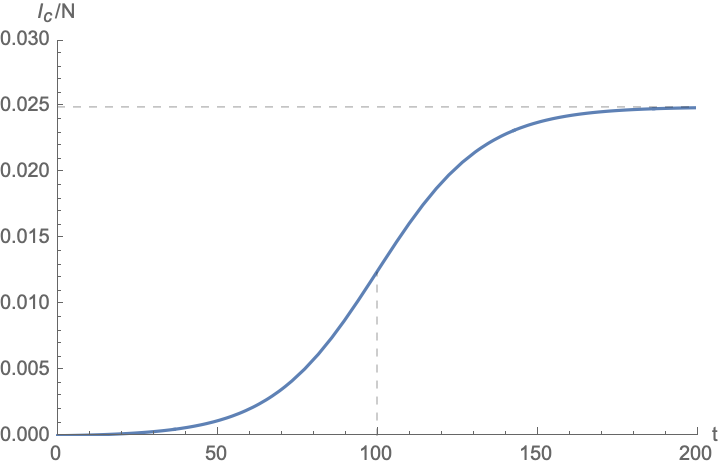}\hspace{1cm}\includegraphics[width=7.5cm]{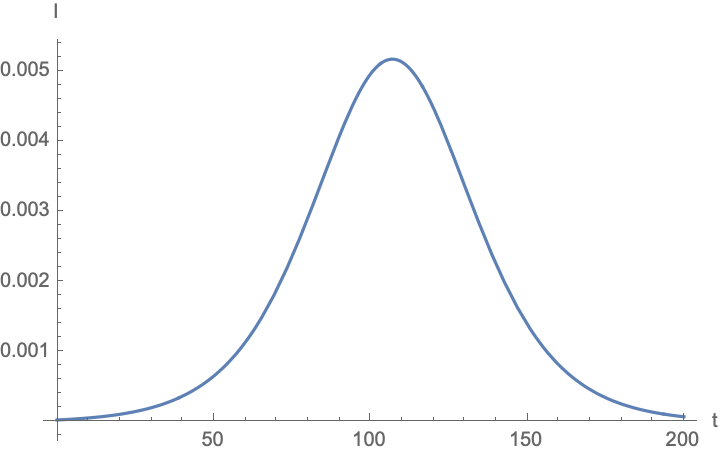}
\end{center}
\caption{Left panel: general form of the logistic function~\ref{LogisticFunction} for $A_0=0.025$, $\lambda_0=0.06$ and $t_0=100$. Right panel: relative number of infected individuals derived from the logistic function for $c=14$. }
\label{Fig:LogisticFunction}
\end{figure}

\begin{figure}[htbp]
\begin{center}
\includegraphics[width=7.5cm]{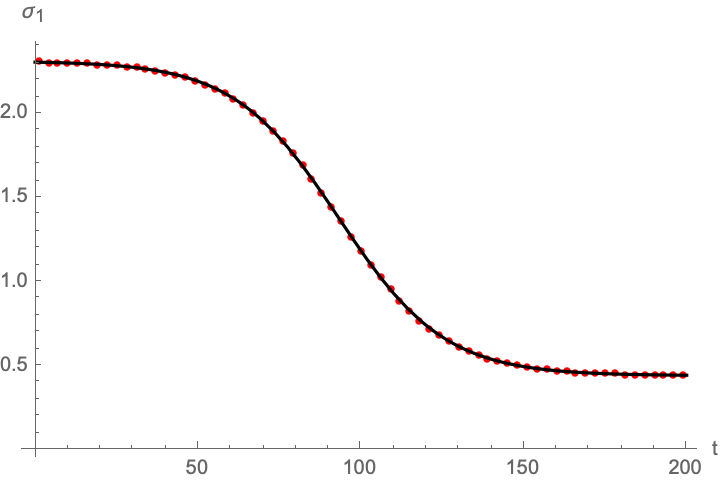}\hspace{1cm}\includegraphics[width=7.5cm]{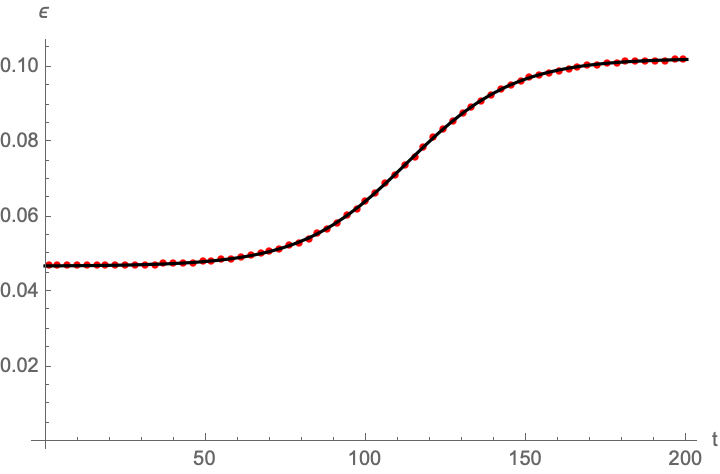}
\end{center}
\caption{Time dependence of the infection rate $\sig_1$ (left panel) and the removal rate $\rhe$ (right panel) needed to reproduce and $\Ic(t)$ of the form (\ref{LogisticFunction}) with the compartmental model (\ref{DiffSIIRV}) (red dots). Both plots use  $A_0=0.025$, $\lambda_0=0.06$, $t_0=100$, $c=14$, $\rvac=0$, $\riv=0$, $\sig_2/\sig_1=1$ and $\mv_0=0$. The interpolating black lines correspond to approximations with logistic functions following (\ref{ApproxTimeDepsGen}) with notably $A_\sig=1.87$, $\delta_\sig=0.44$, $A_\rhe=0.055$ and $\delta_\rhe=0.047$.}
\label{Fig:TimeDepRates}
\end{figure}

\noindent
The schematic form of the function (\ref{LogisticFunction}) is show in the left panel of Figure~\ref{Fig:LogisticFunction}. As shown in \cite{DellaMorte:2020qry} (see also \cite{Cacciapaglia:2021vvu}), in order to obtain the same solution from a compartmental model requires time dependent infection and recovery rates $\rin_{1,2}$ and $\rhe$. In the following we shall study the time dependence of these rates in the SIIRV model (\ref{DiffSIIRV}) that are required to produce as a solution a logistic function of the type (\ref{LogisticFunction}). To this end, however, since $\Ic$ does not distinguish between removed and infectious individuals, we also need to make a (simple) assumption for the (active) relative number of infectious individuals. Indeed, we shall assume that $\mi(t)$ can be extracted by integrating up the growth of $\Ic(t')$ over a period $t'\in[t-c,t]$
\begin{align}
\mi(t)=\int_{t-c}^tdt\,\frac{d\Ic(t')}{dt'}\,dt'=\Ic(t)-\Ic(t-c)\,,\label{ApproxActiveI}
\end{align}
where $c$ is the average amount of time an infectious individual remains infectious.\footnote{We show in appendix~\ref{App:ConstEps} that comparable results can be obtained by assuming that $\rhe$ remains constant throughout the entire outbreak and which does not require (\ref{ApproxActiveI}) to model the number of infectious individuals.} The general form of $\mi$ in (\ref{ApproxActiveI}) is shown in the right panel of Figure~\ref{Fig:LogisticFunction}. Assuming that the ratio $\sig_2/\sig_1$ as well as $\rvac$ remain constant, the time-dependent $(\sig_1,\epsilon)$ to reproduce the same functional dependence of $\Ic(t)$ with the compartmental model in (\ref{DiffSIIRV}), is shown in Figure~\ref{Fig:TimeDepRates}.

Functions of this form were also found in \cite{DellaMorte:2020qry} where the eRG approach was indeed re-interpreted as a time-dependent SIR model. This is indeed expected, since the numerical calculations leading to Figure~\ref{Fig:TimeDepRates} have not taken into account any vaccinations (which we shall consider in the next subsection). Concerning the concrete functional dependence, as is showcased by the black interpolating lines in Figure~\ref{Fig:TimeDepRates}, the time dependence of $(\sig_1(t),\rhe(t))$ can be approximated by logistic functions, concretely
\begin{align}
&\sig_1(t)=A_\sig\left(1-\frac{1}{1+e^{-\lam_\sig(t-t_\sig)}}\right)+\delta_\sig\,,&&\rhe(t)=\frac{A_\rhe}{1+e^{-\lam_\rhe(t-t_\rhe)}}+\delta_{\rhe}\,,\label{ApproxTimeDepsGen}
\end{align}
where 
\begin{align}
&\lam_\sig\sim \lam_\rhe\sim\lam_0\,,&&\text{and}&&t_\sig\sim t_\rhe\sim t_0\,,
\end{align}
while $(A_\sig,\delta_\sig,A_\rhe,\lam_\rhe)$ show a more complicated dependence on $(A_0,\lam_0,t_0)$.
\subsection{Vaccinations}
The approximated time-dependence found in Figure~\ref{Fig:TimeDepRates} assumes the absence of any vaccinations. We therefore next consider a non-vanishing number $\mv_0\neq 0$ of (fully) vaccinated individuals at the outbreak of the wave and a non-trivial vaccination rate $\rvac$. 

\subsubsection{Initial Number of Vaccinated Individuals}
\begin{figure}[htbp]
\begin{center}
\includegraphics[width=7.5cm]{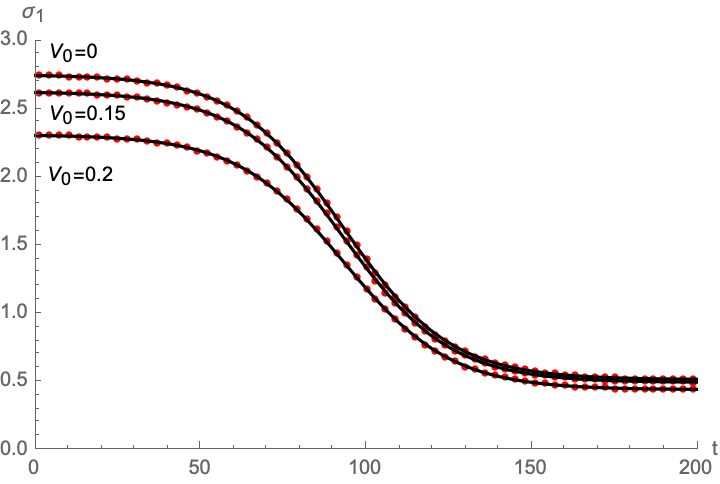}\hspace{1cm}\includegraphics[width=7.5cm]{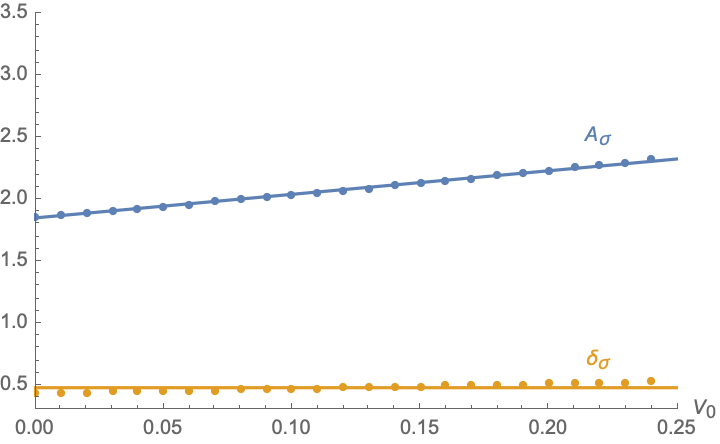}
\end{center}
\caption{Left panel: time dependence of $\sig_1$ for different values of $\mv_0$. Right panel: $A_\sig$ and $\delta_\sig$ as functions of $\mv_0$ required to reproduce (\ref{ModifiedLogisticFunction}) from the compartmental model (\ref{DiffSIIRV}). Both plots use  $A_0=0.025$, $\lambda_0=0.06$, $t_0=100$, $c=14$, $\rvac=0$, $\riv=0$ and $\sig_2/\sig_1=1$.}
\label{Fig:TimeDepRatesV0Dep}
\end{figure}

We first study the impact of $\mv_0$ in the eRG model: in this, we are mainly interested in describing the spread of a disease among a population that is below the herd immunity threshold. In this case, the results of the compartmental SIIRV model in Section~\ref{Sect:HerdSIIRV} suggest that, rather than reproducing (\ref{LogisticFunction}) with parameters $(A_0,\lambda_0,t_0)$, we want to determine $(\sig_1(t),\rhe(t))$ which lead to a solution of (\ref{DiffSIIRV}) of the form 
\begin{align}
\Ic(t)=\frac{A_0(1-\kappa \mv_0)}{1+e^{-\lambda_0(t-t_0)}}\,.\label{ModifiedLogisticFunction}
\end{align}
Here $\kappa$ is a numerical parameter which is close to $\hit$. We also remark that equation (\ref{ModifiedLogisticFunction}) is compatible with results in \cite{cacciapaglia2020us}: there a simple lattice model of susceptible, infectious, removed and vaccinated was considered which showed a linear relation between the asymptotic number of infected individuals and the initial number of vaccinated lattice sites, up to the herd immunity threshold.\footnote{The paper \cite{cacciapaglia2020us} never explicitly discusses herd immunity, but reports a numerical instability of the lattice simulation starting at $V_0\sim 25\%$ of lattice sites. With the data provided for the plot in \cite{cacciapaglia2020us}, this roughly seems compatible with a phase transition at $\hit=1-\frac{\epsilon_*}{\gamma_*}=1/3$, which would indeed manifest itself via a numerical instability in this approach.} Numerically, (\ref{ModifiedLogisticFunction}) requires still a functional dependence of the form (\ref{ApproxTimeDepsGen}) for $(\sig_1(t),\rhe(t))$. Indeed, the parameter $\rhe$ exhibits approximately no dependence on $\mv_0$, while for $\sig_1$ the parameter $A_\sig$ depends linearly on $\mv_0$, while $\delta_\sig$ (and also $\lam_\sig$ and $t_\sig$) to first approximation are independent of $\mv_0$, as shown in Figure~\ref{Fig:TimeDepRatesV0Dep}.

\subsubsection{Vaccination Rate}
The dependence of $\Ic$ on the vaccination rate has previously been discussed in \cite{cacciapaglia2020us} within the eRG formalism. The time dependence of $(\sig_1,\rhe)$ in eq.~(\ref{ApproxTimeDepsGen}) allows us to compare this approach with the compartmental model (\ref{DiffSIIRV}). Concretely, in \cite{cacciapaglia2020us} it has been proposed to supplement 

\begin{wrapfigure}{r}{0.45\textwidth}
${}$\\[-1cm]
\begin{center}
\includegraphics[width=7.5cm]{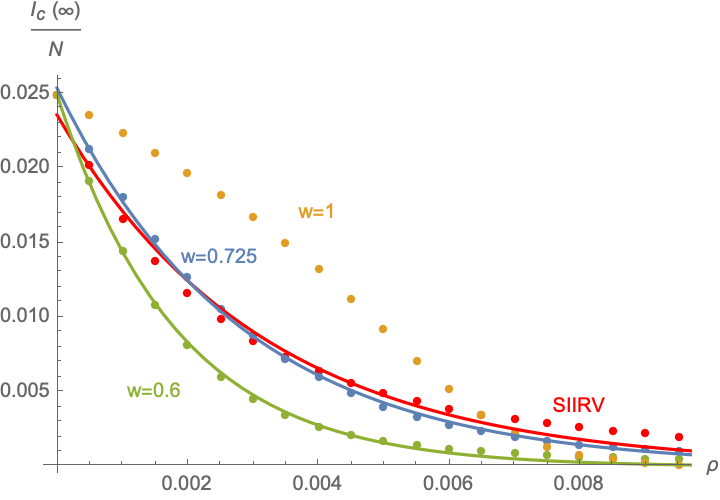}
\caption{ \footnotesize Comparison of the impact of the vaccination rate on the cumulative number of infected individuals predicted by the SIIRV model (\ref{DiffSIIRV}) (with time dependent $(\sig_1,\rhe)$, red curve) to the eRG model using eq.~(\ref{ModifiedVacceRG}) for different values of $w$. The straight lines correspond to interpolations of the form $a\,e^{-b\rho}$ for suitable numerical coefficients $a,b$. The plot uses $A_0=0.025$, $\lambda_0=0.06$, $t_0=100$, $\mv_0=0$ and $\riv=0.2$.}
\label{Fig:Vaccination}
\end{center}
${}$\\[-2.5cm]
\end{wrapfigure}

\noindent
the $\beta$-function (\ref{BetaFunction}) by the following first order differential equations for $\lambda_0$ and~$A_0$ 
\begin{align}
&\frac{d\lambda_0}{dt}=-\rvac \lambda_0(t=0)\,,\nonumber\\
&\frac{dA_0}{dt}=-\rvac\,\left(A_0(t)-\Ic(t)\right)\,.\label{VaccineEquations}
\end{align}
This implies $\lambda_0(t)=\lambda_0(t=0)\left[1-t\,\rvac\right]$, while the equation for the parameter $A_0$ takes into account that in \cite{cacciapaglia2020us} the $\beta$-function was formulated for $\alpha(\Ic)=\ln(\Ic)$ (while we use $\alpha(\Ic)=\Ic$ in this work) and which needs to be solved together with (\ref{BetaFunction}). The second equation in (\ref{VaccineEquations}) was argued for in  \cite{cacciapaglia2020us} by realising that at any given time $t$, the reduction of the asymptotic cumulative number of infected individuals can only depend on $A_0(t)-\Ic(t)$, suggesting a differential equation of the form
\begin{align}
\frac{dA}{dt}=f\left(A_0(t)-\Ic(t)\right)\,,
\end{align}
for a continuous function $f:\mathbb{R}_+\rightarrow\mathbb{R}_+$. The latter was assumed to be linear in \cite{cacciapaglia2020us} (in fact $f(x)=-\rho x$). An immediate generalisation, which allows us to make contact with the time-dependent SIIRV model is the simple modification 
\begin{align}
&\frac{dA_0}{dt}=-\rho\,\left(A_0(t)-\Ic(t)\right)^w\,,&&\text{for} &&w\in[0,1)\,.\label{ModifiedVacceRG}
\end{align}
Indeed, numerical solutions for different choices of $w$ are shown in Figure~\ref{Fig:Vaccination}: while (for the parameters $(A_0,\lambda_0,t_0)$ at $t=0$ chosen in this plot), $w=1$ leads to a qualitatively different function, $w=0.725$ leads to an acceptable agreement. Furthermore,  Figure~\ref{Fig:Vaccination} shows the impact of the initial number of vaccinated individuals on the asymptotic cumulative number of infected individuals, which can be approximated by a function of the form $a\,e^{-b\,\mv_0}$ for some suitable constants $(a,b)$.

\subsection{Health Pass Models}
\begin{figure}[htbp]
\begin{center}
\includegraphics[width=7.5cm]{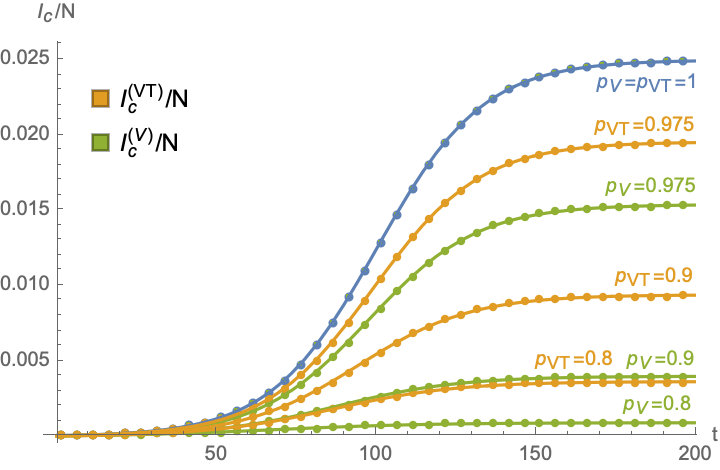}\hspace{1cm}\includegraphics[width=7.5cm]{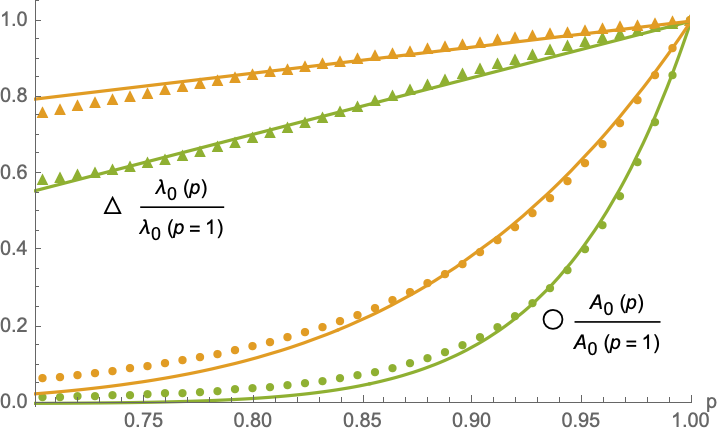}
\end{center}
\caption{Left panel: Cumulative number of infected in the compartmental model (\ref{DiffSIIRVgreenpass}) with time dependent parameters $(\sig_1,\rhe)$ as a function of $\pvt$ and $\pv$ respectively for the two \ep\,models: orange curves represent the model (\ref{DiffSIIRVgreenpass3G}) and green curves the model (\ref{DiffSIIRVgreenpass}). The blue curve (with $\pvt=\pv=1$) is identical in both models (and corresponds to the case of no \ep). All parameters (including the time dependence of $(\sig_1,\rhe)$) are as in Figure~\ref{Fig:TimeDepRates}. Right panel: $p$-dependence of the parameters $(A_0,\lambda_0)$ relative to the case $p=1$: circles represent numerical values of $\frac{A_0(p)}{A_0(p=1)}$ while triangles represent numerical values of $\frac{\lambda_0(p)}{\lambda_0(p=1)}$, with orange symbols computed using the model (\ref{DiffSIIRVgreenpass3G}) and green symbols correspond to the model (\ref{DiffSIIRVgreenpass3G}). The solid lines represent interpolations of the numerical solutions with an exponential function of the form (\ref{SolpImp}) for $\frac{A_0(p)}{A_0(p=1)}$ and a linear function for $\frac{\lambda_0(p)}{\lambda_0(p=1)}$. The plots use $A_0=0.025$, $\lambda_0=0.06$ (at $t=0$), $t_0=100$, $\mv_0=0.3$ and $\riv=0.15$.}
\label{Fig:ApproxLogGreenPassComp}
\end{figure}

As a next step, we assume that the general form of a time dependent $(\sig_1,\epsilon)$ remains valid also after implementing either of the two \ep\, models (\ref{DiffSIIRVgreenpass3G}) or (\ref{DiffSIIRVgreenpass}) and the only modification is due to the (constant) parameter $\pvt$ and $\pv$ respectively. In this case, numerical solutions indicate that the cumulative number of infected individuals can still be well approximated by a logistic function (\ref{LogisticFunction}) (see the left panel of Figure~\ref{Fig:ApproxLogGreenPassComp}), albeit with $p$-dependent parameters $(A_0,\lambda_0,t_0)$ as shown in the right panel of Figure~\ref{Fig:ApproxLogGreenPassComp}. While for small values of $1-p$ (\emph{i.e.} for $p\in[0.7,1]$)$\lambda(p)$ 

\noindent
and $\lambda(p)\,t_0(p)$ can be interpolated by linear functions in $p$, the asymptotic number of infected individuals is approximated by an exponential function 
\begin{align}
&A_0(p)\sim A_0(p=1)\,\text{exp}\left(\theta\,\frac{p-1}{p}\right)\,,\label{SolpImp}
\end{align}
with $\theta\in\mathbb{R}_+$. The approximation (\ref{SolpImp}) can in fact be used for the entire range of $p\in[0,1]$. The constant $\theta$ implicitly depends on the remaining parameters of the problem (notably $\rvac$, $\mv_0$ and $A_0(p=1)$). Furthermore, we also show the asymptotic cumulative number of infected in-

\begin{wrapfigure}{r}{0.52\textwidth}
\begin{center}
\includegraphics[width=9cm]{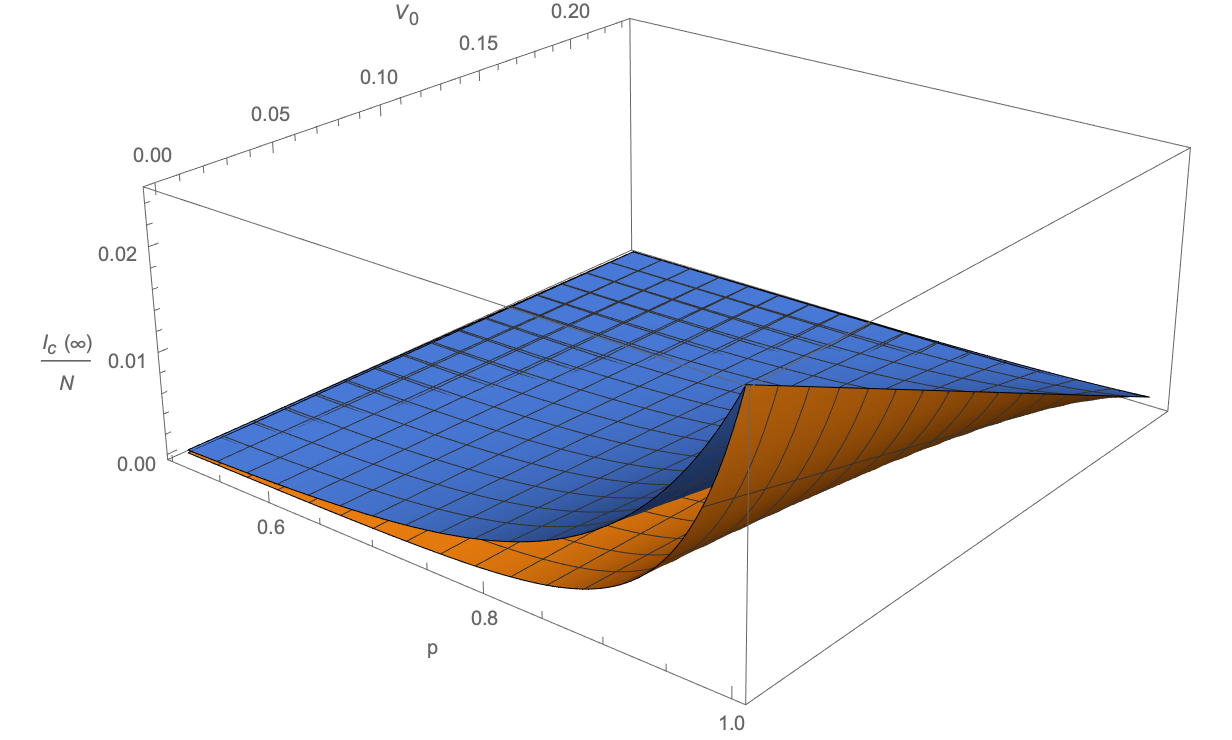}
\caption{\footnotesize Plot of $\Ic(\infty)/N$ as a function of $(p,V_0)$. The blue surface refers to a \vtep-model, while the orange one refers to a \vep-model. The plots use $\rvac=0.0008$, $\riv=0.2$ and $\sig_2=\sig_1$ as well as $A_0(p=1)=0.025$, $t_0(p=1)=100$ and $\lambda_0(p=1)=0.06$. }
\label{Fig:PlanePlotpV0}
\end{center}
${}$\\[-1cm]
\end{wrapfigure}
\begin{figure}[hbtp]
\begin{center}
\includegraphics[width=7.5cm]{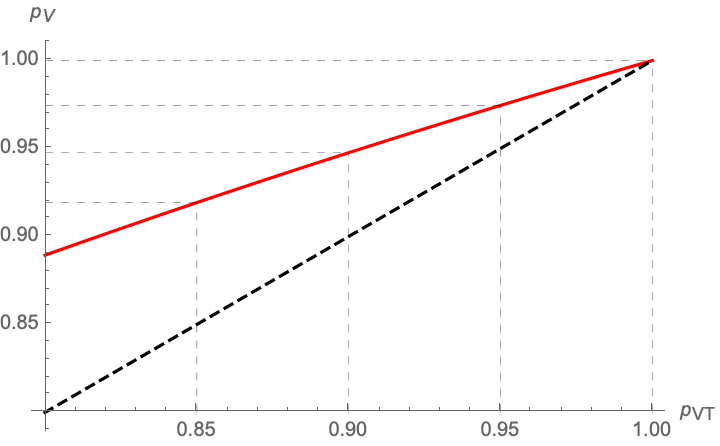}\hspace{1cm}\includegraphics[width=7.5cm]{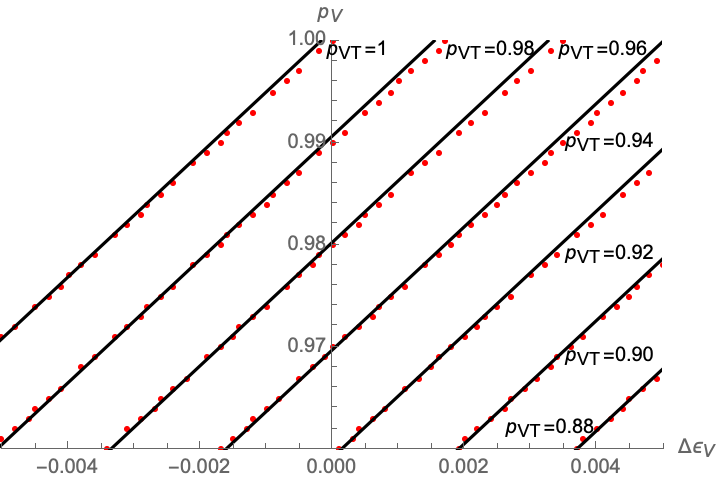}
\end{center}
\caption{Imposing the same asymptotic cumulative number of infected individuals leads to implicit relations between the parameters of the Green Pass models (\ref{DiffSIIRVgreenpass3G}) and (\ref{DiffSIIRVgreenpass}). Left panel: assuming all other parameters of the models to remain the same entails a linear equivalence relation between the efficacy $(\pv,\pvt)$ of the models (\ref{DiffSIIRVgreenpass3G}) and (\ref{DiffSIIRVgreenpass}). Right panel: assuming the change (\ref{DeltaShift}) in the time-dependent parameter $\rhe_{\text{V}}(t)$ leads to a degeneracy of the parameters that lead to the same asymptotic cumulative number of infected individuals for the model (\ref{DiffSIIRVgreenpass}): all combinations of parameters along the (interpolating) solid lines are 'equivalent' to model (\ref{DiffSIIRVgreenpass3G}) with the indicated value of $\pvt$. The plots use $A_0=0.025$, $\lambda_0=0.06$ (at $t=0$), $t_0=100$, $\mv_0=0.3$ and $\riv=0.15$.}
\label{Fig:ConversionMod31}
\end{figure}


\noindent
dividuals as a function of $(p,\mv_0)$ in Figure~\ref{Fig:PlanePlotpV0}. As in the case of constant rates (see Section~\ref{Sect:GreenPassesConstantRates}) we can consider models of the type (\ref{DiffSIIRVgreenpass3G}) and (\ref{DiffSIIRVgreenpass}) as equivalent if they lead to the same asymptotic cumulative number of infected individuals, \emph{i.e.} if the same number of people got infected during the entire duration of the wave. As before, imposing this condition leads to an equivalence in the parameters of both models. Indeed, the left panel of Figure~\ref{Fig:ConversionMod31} shows this equivalence between efficacies $\pvt$ and $\pv$, which we find to still be roughly linearly related, exactly as in the case of constant infection and removal rates (see Figure~\ref{Fig:ComparisonGreenPassEquiv}). However, for the current models with time dependent rates, we have more possibilities to change the removal rate. For example, considering the time dependence (\ref{ApproxTimeDepsGen}) of $\rhe_1(t)$, we can consider
\begin{align}
\delta_{\rhe}\rightarrow \delta_{\rhe}+\Delta_{\rhe}\,,\label{DeltaShift}
\end{align}
for a constant parameter $\Delta_{\rhe}$. Comparing numerical computations of different values of $(\pv,\Delta_{\rhe_1})$ leads to equivalences as shown in the right panel of Figure~\ref{Fig:ConversionMod31}.

\section{Examples}\label{Sect:Examples}
In the following we apply the SIIRV model with time dependent infection and recovery rates developed in Section~\ref{Sect:CompartmentalSIIRV}, which is equivalent to an eRG model with vaccinations discussed in Section~\ref{Sect:EPeRG}, as well as its generalisations that include \ep\, to the situations of certain countries in late-summer of 2021. Indeed, during this period many countries are threatened by a potential new wave of COVID-19 caused by the Delta variant of \cov, in spite of a vaccination campaign.

\subsection{Germany}
\subsubsection{Data}
We first consider the situation in Germany: since the outbreak of the COVID-19 pandemic, Germany has been hit by three waves, which can be seen from the cumulative number of infected individuals as plotted in Figure~\ref{Fig:FittingWavesGermany}. We have fitted each wave individually by a logistic function
\begin{align}
\Ic^{\text{wave}}(t)=\Icn{0}+\frac{A_0}{1+e^{-\lambda_0(t-t_0)}}\,,\label{IcWaveLogistic}
\end{align}
where the parameters for each wave are given in the following table:
\begin{figure}[htbp]
\begin{center}
\includegraphics[width=12cm]{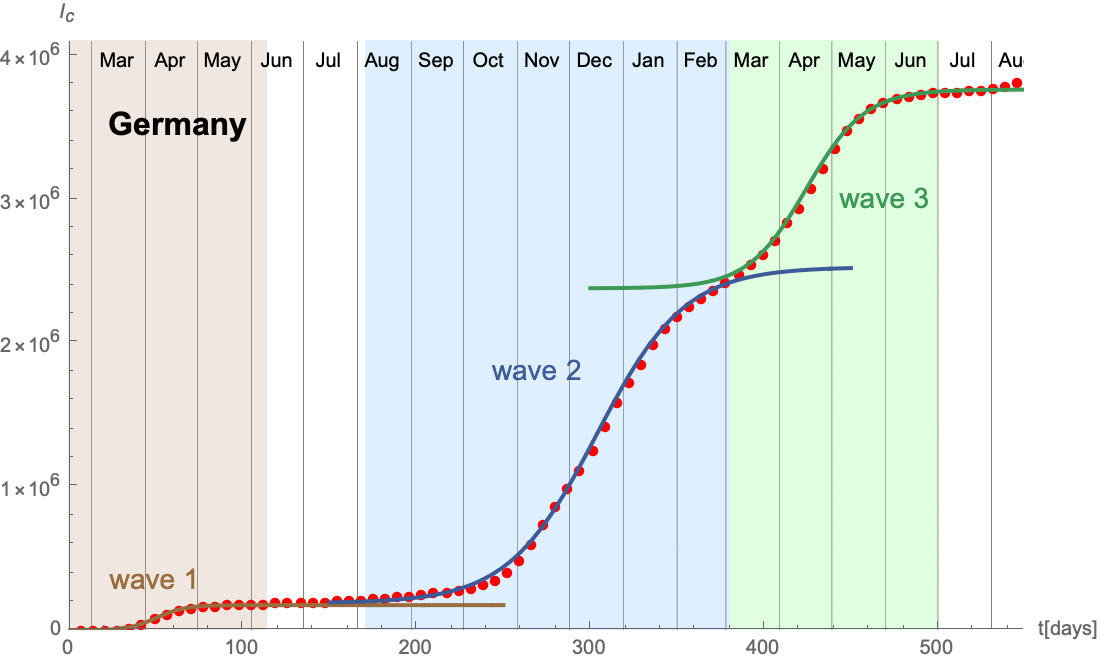}
\end{center}
\caption{Fitting the first three waves in Germany. The red dots represent the cumulative number of infected individuals as reported in \cite{Worldometer}. The coloured regions indicate the time frames used to fit the three different waves, whose fits are shown by the solid lines. The fit parameters are exhibited in the table in the text.}
\label{Fig:FittingWavesGermany}
\end{figure}

\begin{figure}[htbp]
\begin{center}
\includegraphics[width=7.5cm]{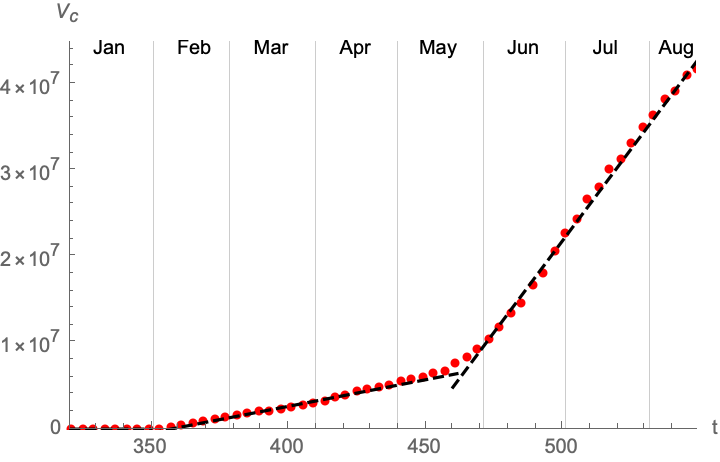}\hspace{1cm}\includegraphics[width=7.5cm]{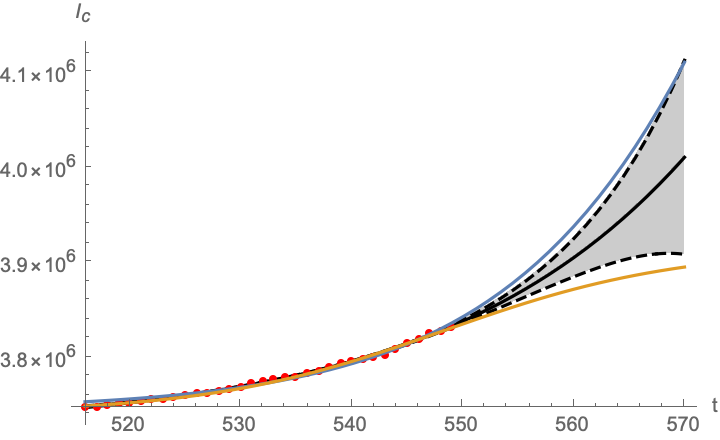}
\end{center}
\caption{Left panel: Cumulative number of fully immunised individuals in Germany as a function of time. Right panel: the cumulative number of cases suggests the onset of a fourth wave in Germany in the late summer/beginning of fall. The gray region indicates the 90\% confidence interval of the fit of the reported cases with a logistic function, with the blue and orange curve representing an approximation of the extremal cases through logistic functions.}
\label{Fig:FittingVaccinesGermany}
\end{figure}

\begin{center}
\begin{tabular}{|c||c|c|c|c|}\hline
&&&&\\[-10pt]
{\bf wave} & $A_0$ & $\lambda_0$ & $t_0$ & $\Icn{0}$\\[2pt]\hline\hline
&&&&\\[-10pt]
wave 1 & $177112\pm 918$ & $0.119\pm 0.003$ & $50.2\pm 0.3$ & 0\\[2pt]\hline
&&&&\\[-10pt]
wave 2 & $2.33692\cdot 10^6\pm 10080$ & $0.0395\pm 0.0003$ & $303.4\pm 0.2$ & $190913\pm 3631$\\[2pt]\hline
&&&&\\[-10pt]
wave 3 & $1.38258\cdot 10^6\pm 10061$ & $0.0595\pm 0.0008$ & $424.3\pm 0.2$ & $2.37996\cdot 10^6\pm 7943$\\[2pt]\hline
\end{tabular}
\end{center}

\noindent
The cumulative number of fully immunised individuals\footnote{Following a recommendation \cite{RKI} of the Robert Koch Institut (RKI), we count an individuals as fully immunised 14 days after having received the last dose of a full vaccination scheme.} is shown in Figure~\ref{Fig:FittingVaccinesGermany}, which can be approximated by a piecewise linear function: indeed, a strong increase in the vaccination rate can be seen in the middle of May 2021. Finally, analysing the data of the month of July 2021 suggests the onset of a fourth wave. Fitting of these data with a logistic function is shown in the right panel of Figure~\ref{Fig:FittingWavesGermany}, which, however, also indicates the $90\%$-confidence interval for the development of the following 3 weeks. 

In the following, we shall first analyse the past waves 2 and 3, to show the reconciliation of the time-dependent SIIRV model (\ref{DiffSIIRV}) developed in the previous sections with the eRG approach. In a second step, we propose extremal parameter sets for the compartmental model based on the (incomplete) fitting of the eRG model for the impending fourth wave, in order to obtain estimations for the development in late summer/early fall of 2021 regarding the implementation of a \ep.

\subsubsection{Time Dependent SIIRV Model for Waves 2 and 3}
\begin{figure}[htbp]
\begin{center}
\includegraphics[width=7.5cm]{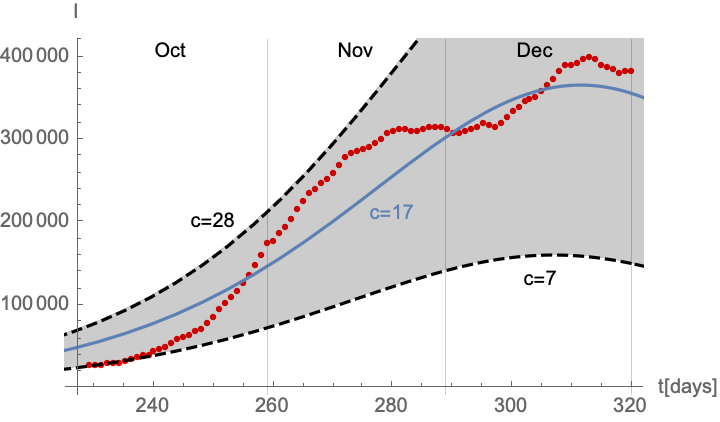}\hspace{1cm}\includegraphics[width=7.5cm]{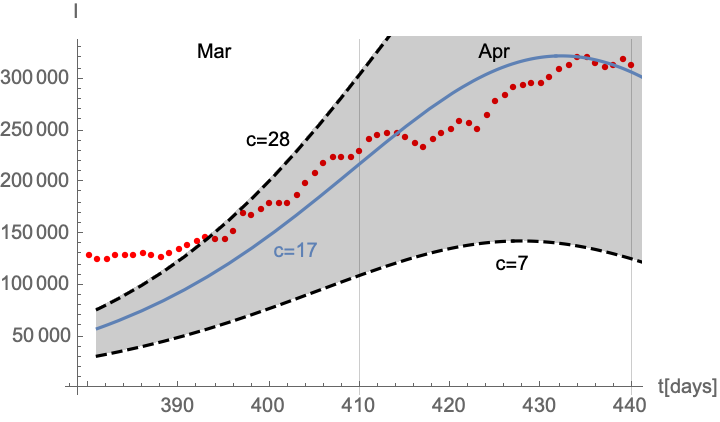}
\end{center}
\caption{Fitting of the (active) number of infectious individuals derived from the cumulative number of infected individuals (black band) compared to the reported cases (red dots). The approximation shows the range of the parameter $c\in[7,28]$, with the best fit indicated by the blue line for $c=17$.}
\label{Fig:FittingActivesGermany}
\end{figure}

First, using (\ref{ApproxActiveI}), we compute the number of (active) infectious individuals associate with the cumulative number of infected individuals for waves 2 and 3. As shown in Figure~\ref{Fig:FittingActivesGermany}, comparing with the active cases reported in \cite{Worldometer} is best accommodated using $c=17$ days.

Based on the functions $\mi(t)$ and $\Ic$, we next derive time-dependent parameters $(\sig_1(t),\rhe(t))$ such that the SIIRV model (\ref{DiffSIIRV}) reproduces the cumulative number of infected individuals. For both waves, we find that the rates can be approximated by logistic functions of the form given in eq.~(\ref{ApproxTimeDepsGen}), where the fitting parameters $(A_{\sig,\rhe},\lambda_{\sig,\rhe},\tau_{\sig,\rhe},\delta_{\sig,\rhe})$ are shown in the following table:

 \begin{center}
\begin{tabular}{|c|c|c|c|c|c|c|c|c|}\hline
& \multicolumn{2}{c|}{$A_{\sig,\rhe}$} & \multicolumn{2}{c|}{$\lambda_{\sig,\rhe}$} & \multicolumn{2}{c|}{$t_{\sig,\rhe}$} & \multicolumn{2}{c|}{$\delta_{\sig,\rhe}$}\\[2pt]\hline
&&&&&&&&\\[-12pt]
& wave 2 & wave 3 & wave 2 & wave 3 & wave 2 & wave 3 & wave 2 & wave 3\\[2pt]\hline\hline
&&&&&&&&\\[-12pt]
$\sig_1$ & 1.333 & 2.156 & 0.0403 & 0.064 & 124.8 & 36.9 & 0.547 & 0.43\\[2pt]\hline
&&&&&&&&\\[-12pt]
$\rhe$ & 0.037 & 0.056 & 0.0395 & 0.0595 & 147.5 & 58.3 & 0.046 & 0.039\\[2pt]\hline
\end{tabular}
\end{center}

\begin{figure}[htbp]
\begin{center}
\includegraphics[width=7.5cm]{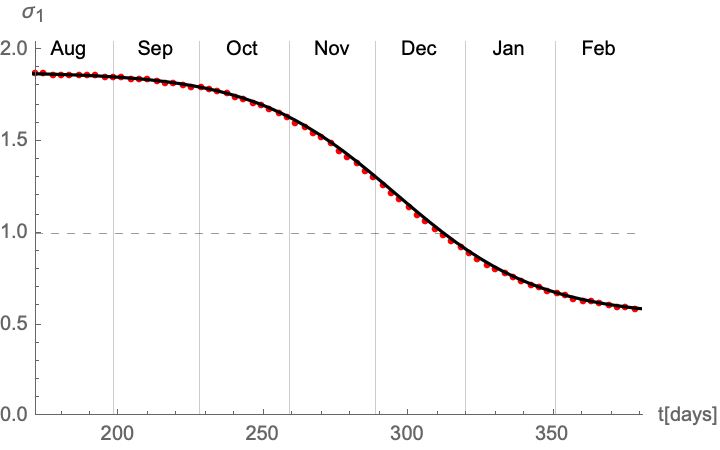}\hspace{1cm}\includegraphics[width=7.5cm]{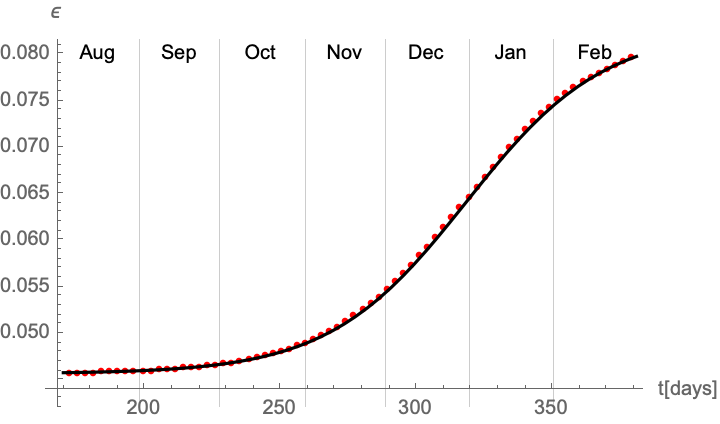}\\
\includegraphics[width=7.5cm]{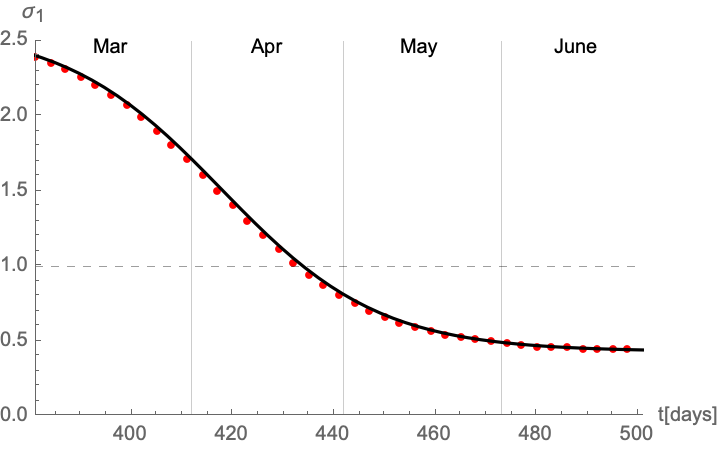}\hspace{1cm}\includegraphics[width=7.5cm]{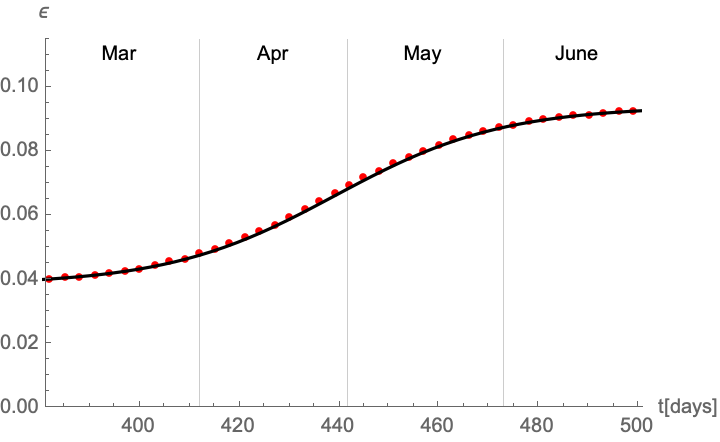}\\
\end{center}
\caption{The time-dependent parameters $(\sig_1,\rhe)$ needed to match the compartmental model (\ref{DiffSIIRV}) (red dots) along with their interpolations (black lines) following (\ref{ApproxTimeDepsGen}). Top panels: wave 2, bottom panels: wave 3. In the latter case we have used $\riv=0.15$ as well as $\rhe=0.0008$.}
\label{Fig:ComparisonFitRealActives}
\end{figure}

\begin{wrapfigure}{r}{0.5\textwidth}
\begin{center}
${}$\\[-1.5cm]
\includegraphics[width=7.5cm]{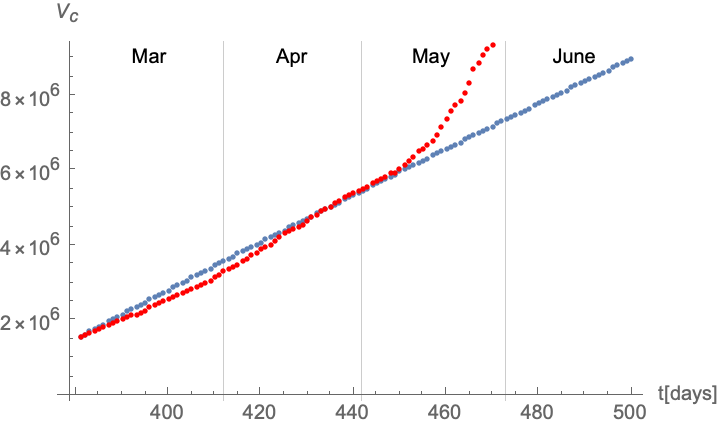}
\caption{\footnotesize Comparison of cumulative number of fully vaccinated individuals reported by the RKI \cite{RKI} (red) with the numerical calculation of the SIIRV model (blue).}
\label{Fig:GermanyVaccComp}
\end{center}
${}$\\[-2cm]
\end{wrapfigure}

\noindent
The functions $(\sig_1,\rhe)$ are also plotted in Figure~\ref{Fig:ComparisonFitRealActives}. During wave 2, vaccines were not available, such that the above fitting uses $\riv=0=\rvac$, while for wave 3 we have used $\riv=0.15$ as well as $\rvac=0.0008$. In Figure~\ref{Fig:GermanyVaccComp}, the cumulative number of fully vaccinated individuals is compared with the actual number reported by the RKI \cite{RKI}: while the agreement is excellent for the months March and April, the vaccination rate has been significantly increased in Germany in the middle of May.

\noindent
  Since this occurred towards the end of wave three (\emph{i.e.} after the maximum in the number of active infectious individuals has been reached, as can be seen in the right panel of Figure~\ref{Fig:FittingActivesGermany}), we have ignored this effect and have fitted the entire wave with $\rho=0.0008$.

\subsubsection{Wave 4}
\begin{figure}[htbp]
\begin{center}
\includegraphics[width=7.5cm]{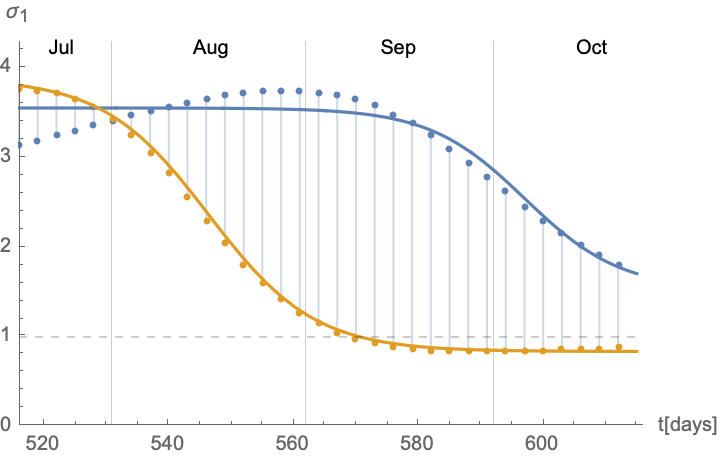}\hspace{1cm}\includegraphics[width=7.5cm]{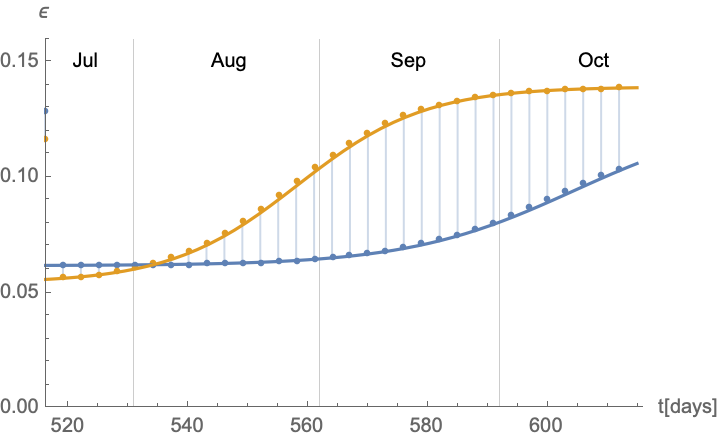}\\
\end{center}
\caption{Time dependent parameters $\sig_1$ (left panel) and $\rhe$ (right panel) for the extremal cases of wave 4. The blue and orange colours are correlated with the curves in the right panel of Figure~\ref{Fig:FittingWavesGermany}.}
\label{Fig:ComparisonFitWave4}
\end{figure}

The final step consists in applying time-dependent parameters $(\sig_1,\rhe)$ to the \ep\, model (\ref{DiffSIIRVgreenpass}) to model and predict the impending fourth wave. We consider as two extremal cases logistic functions obtained from fits for the currently available data up to mid August as shown in Figure~\ref{Fig:ComparisonFitWave4}. In this way, we obtain a 'band' for the cumulative number of infected individuals. The parameters for the logistic functions (represented by the blue and orange curve visible in Figure~\ref{Fig:ComparisonFitWave4}) are respectively given by
\begin{align}
&A_0^+=2.5\cdot 10^6\,,&&\lam_0^+=0.072\,,&&t_0^+=594.5\,,&&\delta_0^+=3.74\cdot 10^6\,,\nonumber\\
&A_0^-=174533\,,&&\lam_0^-=0.094\,,&&t_0^-=548.3\,,&&\delta_0^-=3.74\cdot 10^6\,.
\end{align}

\noindent
Based on these values, we can develop a time-dependent SIIRV model, with time dependent parameters $(\sig_1,\epsilon)$, which are shown in Figure~\ref{Fig:ComparisonFitWave4}. These curves follow the general form of eq.~(\ref{ApproxTimeDepsGen}) with the parameters

\begin{center}
\begin{tabular}{|c||c|c||c|c||c|c||c|c|}\hline
&&&&&&&&\\[-12pt]
 & $A_{\sig,\rhe}^+$ & $A_{\sig,\rhe}^-$ & $\lambda_{\sig,\rhe}^+$ & $\lambda_{\sig,\rhe}^-$ & $t_{\sig,\rhe}^+$ & $t_{\sig,\rhe}^-$ & $\delta_{\sig,\rhe}^+$ & $\delta_{\sig,\rhe}^-$\\[2pt]\hline\hline
&&&&&&&&\\[-12pt]
 $\sig_1$ &  2.024 & 3.047 & 0.131 & 0.119 & 81.1 & 30.4 & 1.529 & 0.838 \\[2pt]\hline
 &&&&&&&&\\[-12pt]
 $\rhe$ &  0.065 & 0.085 & 0.072 & 0.094 & 88.5 & 42.3 & 0.062 & 0.054 \\[2pt]\hline
\end{tabular}
\end{center}

\begin{figure}[htbp]
\begin{center}
\includegraphics[width=7.5cm]{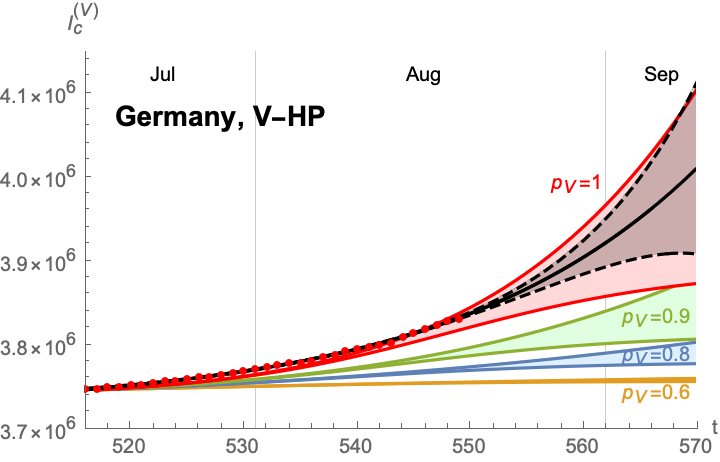}\hspace{1cm}\includegraphics[width=7.5cm]{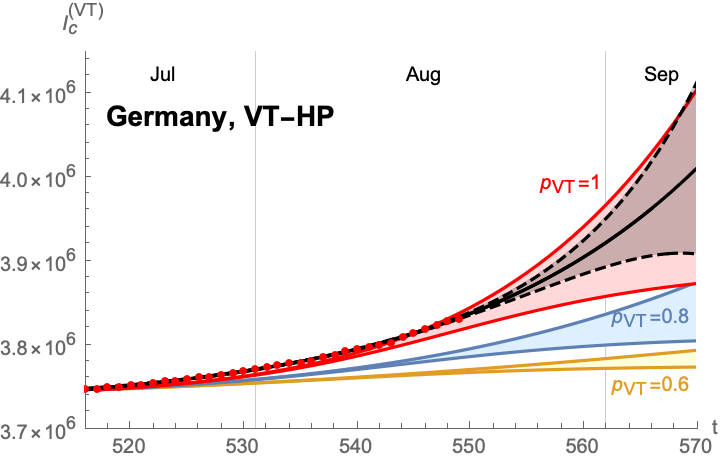}
\end{center}
\caption{Time evolution of the cumulative number of infected individuals for different values of the efficacy of a \vep\, (right panel) or a \vtep\,(right panel). We assume that the model has been introduced on 07/07/2021. Both cases use $\riv=0.15$ (based on an average of the efficacy of each vaccine weighted by the distribution among the population), $\rvac=0.008$ and $\sig_2/\sig_1=1$.}
\label{Fig:GreenPassGermany}
\end{figure}

\begin{figure}[htbp]
\begin{center}
\includegraphics[width=7.5cm]{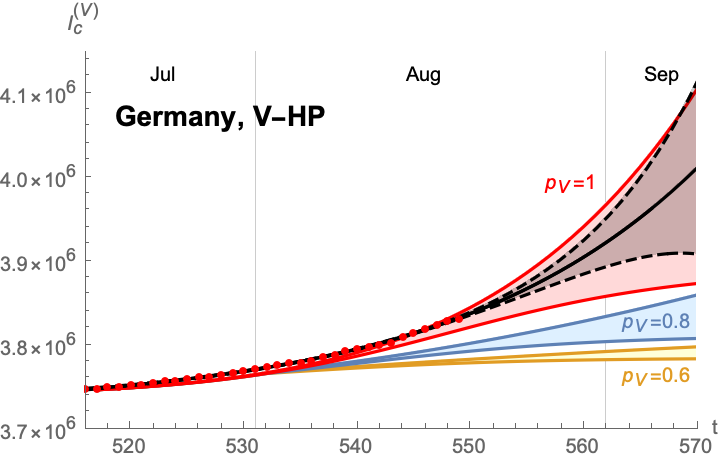}\hspace{1cm}\includegraphics[width=7.5cm]{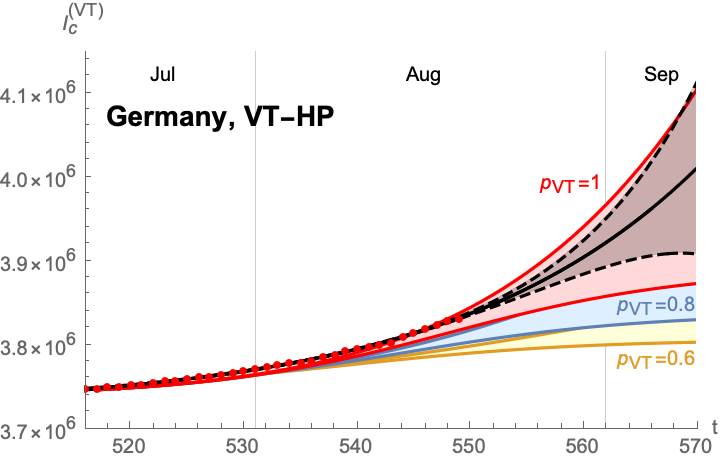}
\end{center}
\caption{ime evolution of the cumulative number of infected individuals for different values of the efficacy of a \vep\, (right panel) or a \vtep\,(right panel). We assume that the model has been introduced on 01/08/2021. Both cases use $\riv=0.15$ (based on an average of the efficacy of each vaccine weighted by the distribution among the population), $\rvac=0.008$ and $\sig_2/\sig_1=1$.}
\label{Fig:GreenPassGermanyLater}
\end{figure}

\begin{figure}[htbp]
\begin{center}
${}$\\[-0.5cm]
\includegraphics[width=7.5cm]{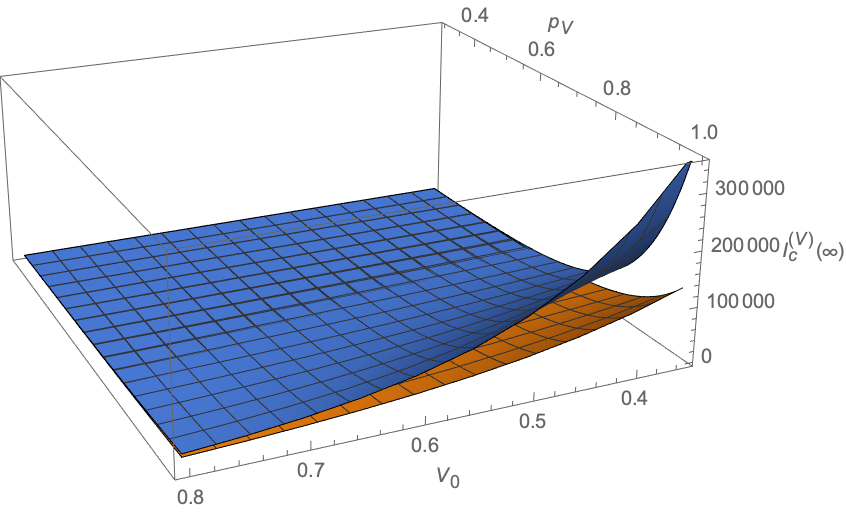}\hspace{1cm}\includegraphics[width=7.5cm]{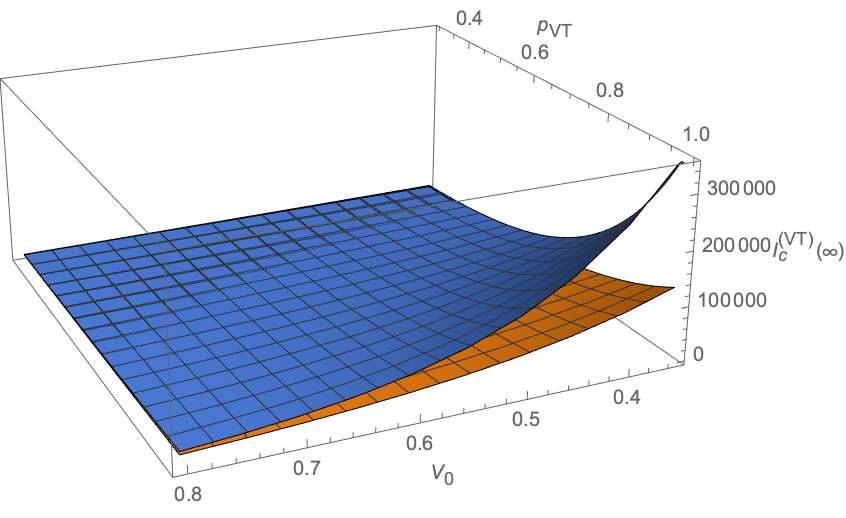}
\caption{Extremal predictions of $\Ic$ in the plane of $(p,\mv_0)$, assuming $\rvac=0.008$, $\riv=0.15$. The left panel shows the plot for a \vep, while the right panel for a \vtep.}
\label{Fig:GreenPassGermanyPlane}
\end{center}
${}$\\[-1.5cm]
\end{figure}

\noindent
Finally, implementing the Green pass model based on these extremal cases is shown in Figure~\ref{Fig:GreenPassGermany}. Starting from the band of cumulative number of infected individuals (for $p=1$), we obtain a similar band for each $p<1$. The plots for a \vtep\, and \vep\, respectively are shown for $\sig_2/\sig_1=1$ in Figure~\ref{Fig:GreenPassGermany}, assuming that the Green Pass has been introduced on 07/July/2021. Figure~\ref{Fig:GreenPassGermanyLater} shows the same analysis assuming that the \ep\, had been introduced 01/08/2021.The plots in Figure~\ref{Fig:GreenPassGermany} suggest that a new wave in Germany could be stopped by reducing the contacts among non-vaccinated individuals by roughly 20-40\%.
 
\begin{figure}[htbp]
\begin{center}
\includegraphics[width=7.5cm]{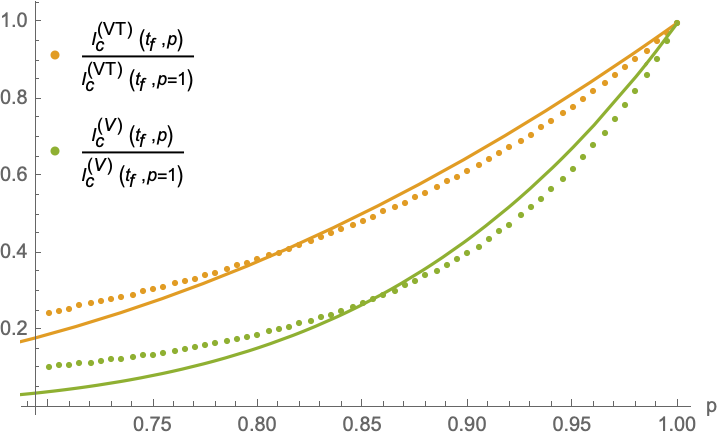}\hspace{1cm}
\includegraphics[width=7.5cm]{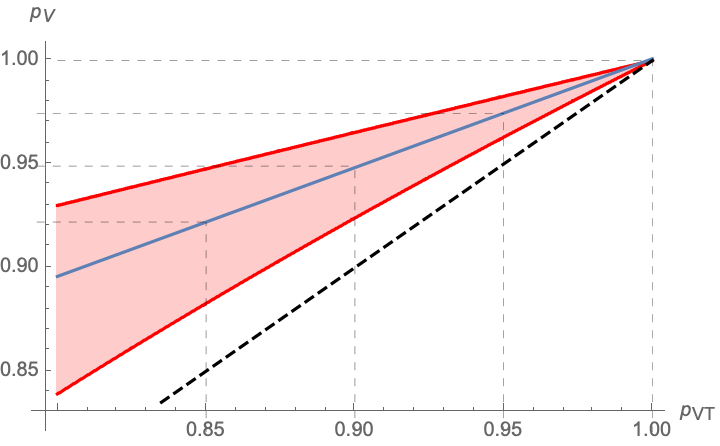}
\end{center}
\caption{Comparison of the \vep\, model (\ref{DiffSIIRVgreenpass}) and the \vtep\, model(\ref{DiffSIIRVgreenpass3G}): The left panel shows the cumulative number of infected individuals at $t_f=15/09/2021$ as a function of $p$ (normalised to the value of $p=1$). The orange curve corresponds to the model (\ref{DiffSIIRVgreenpass}) and the green curve to the one in (\ref{DiffSIIRVgreenpass3G}). The right panel shows the equivalence for the parameters $\pvt$ and $\pv$ of these two models, taking into account the incertitude inherent in the approximations: the red band indicates equivalent values of these parameters that lead to the same value of $\Ic(t_f)$ with the blue line corresponding to equivalence obtained comparing equivalent extrapolations of the data in each case.}
\label{Fig:GermanyComparisonGreenPassEquiv}
\end{figure}

Finally, Figure~\ref{Fig:GreenPassGermanyPlane} shows the cumulative number of infected in the $(p,\mv_0)$-plane. Furthermore, we have compared the efficacy of the \vtep\, and \vep\, in the case of Germany in Figure~\ref{Fig:GermanyComparisonGreenPassEquiv}: the left panel shows the (normalised) cumulative number of infected individuals at $t_f=15/09/2021$\footnote{Rather than the asymptotic number of infected individuals at the end of the wave, we have chosen a date roughly a months after the last available data for the comparison.}, along with an approximation of the form (\ref{SolpImp}). The right panel shows which values of $\pv$ and $\pvt$ lead to the same cumulative number of infected individuals at $t_f$: the red band corresponds to the uncertainty related to the two extremal cases we have developed to extrapolate the data. In fact, the extrema of this band arise when comparing the most optimistic extrapolation for the \vep\, with the worst case approximation of \vtep\, (and vice versa). The blue line corresponds to a comparison of equivalent extrapolations and suggests roughly
\begin{align}
2(1-\pv)\sim 1-\pvt\,.\label{EquivRelPvPvt}
\end{align}
This means, assuming that all other parameters remain roughly the same, the reduction in the contacts in the \vtep\, needs to be roughly twice as large as in the \vep-model to achieve the same cumulative number of infected.

\subsection{Austria}
\subsubsection{Data and Previous Waves}
\begin{figure}[htbp]
\begin{center}
\includegraphics[width=12cm]{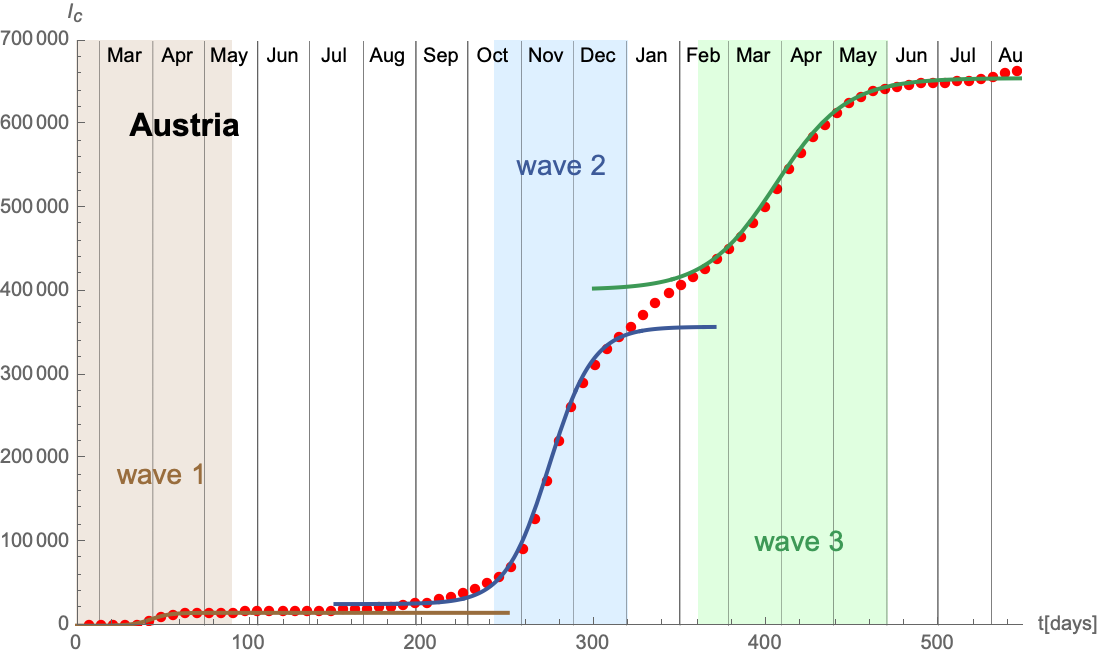}
\end{center}
\caption{Fitting the first Three Waves in Austria. The red dots represent the cumulative number of infected individuals as reported on \cite{Worldometer}. The coloured regions indicate the time frames used to fit the three different waves, whose fits are shown by the solid lines. The fit parameters are exhibited in the table in the text.}
\label{Fig:FittingWavesAustria}
\end{figure}

We next consider the situation in Austria: similarly to Germany, Austria has also been hit by three waves since the beginning of the COVID-19 pandemic, and is confronted with growing infection numbers in the late summer/early fall of 2021, indicating a potential upcoming fourth wave. The cumulative number of infected individuals since the beginning of the pandemic is shown in Figure~\ref{Fig:FittingWavesAustria}: as can be seen, each wave can be fitted with a logistic function of the form (\ref{IcWaveLogistic}) with the following numerical parameters

\begin{center}
\begin{tabular}{|c||c|c|c|c|}\hline
&&&&\\[-10pt]
{\bf wave} & $A_0$ & $\lambda_0$ & $t_0$ & $\Icn{0}$\\[2pt]\hline\hline
&&&&\\[-10pt]
wave 1 & $15361\pm 70$ & $0.182\pm 0.004$ & $43.0\pm 0.2$ & 0\\[2pt]\hline
&&&&\\[-10pt]
wave 2 & $331987\pm 4907$ & $0.0785\pm 0.002$ & $274.0\pm 0.3$ & $25832.6\pm 3512$\\[2pt]\hline
&&&&\\[-10pt]
wave 3 & $253295\pm 1031$ & $0.0491\pm 0.0003$ & $405.6\pm 0.1$ & $402220\pm 722$\\[2pt]\hline
\end{tabular}
\end{center}

\noindent
The number of (active) infectious individuals derived for waves 2 and 3 from these data is 
\begin{figure}[htbp]
\begin{center}
\includegraphics[width=7.5cm]{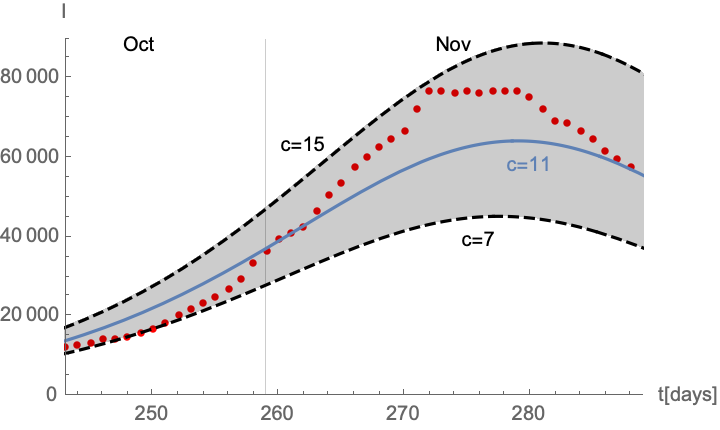}\hspace{1cm}\includegraphics[width=7.5cm]{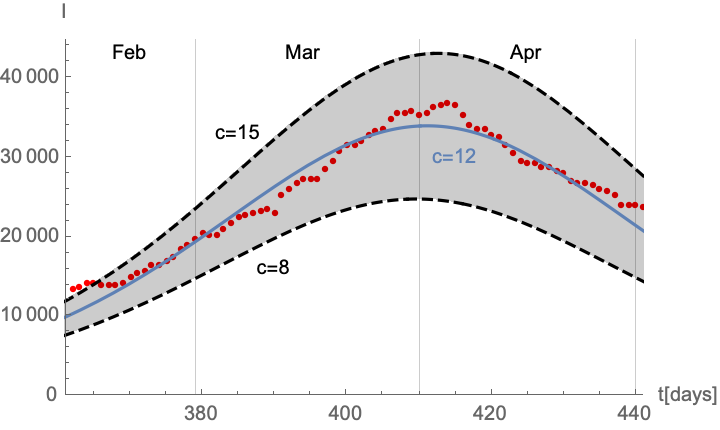}
\end{center}
\caption{Fitting of the (active) number of infectious individuals derived from the cumulative number of infected individuals (black lines) compared to the reported cases (red dots). The gray band represents different values for the parameter $c$ in (\ref{ApproxActiveI}) with the blue line giving the best numerical fit.}
\label{Fig:FittingActivesAustria}
\end{figure}


\noindent
shown in Figure~\ref{Fig:FittingActivesAustria} and compared to the cases reported on \cite{Worldometer}. The grey band shows different approximations depending on the parameter $c$, which in turn is a measure for how long on average an infected individual remains infectious. Similar to the situation in Germany (see Figure~\ref{Fig:FittingActivesGermany}) good agreement can be achieved for $c\in[7,15]$ with the best fit corresponding to $c=11$ and $c=12$ for waves 2 and 3 respectively.

\begin{figure}[htbp]
\begin{center}
\includegraphics[width=7.5cm]{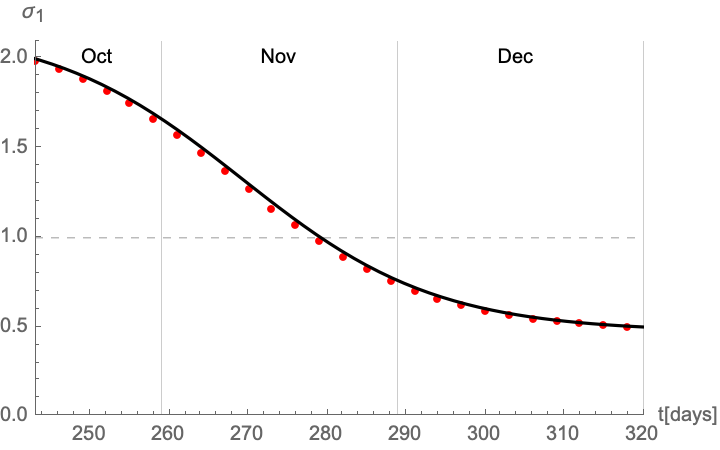}\hspace{1cm}\includegraphics[width=7.5cm]{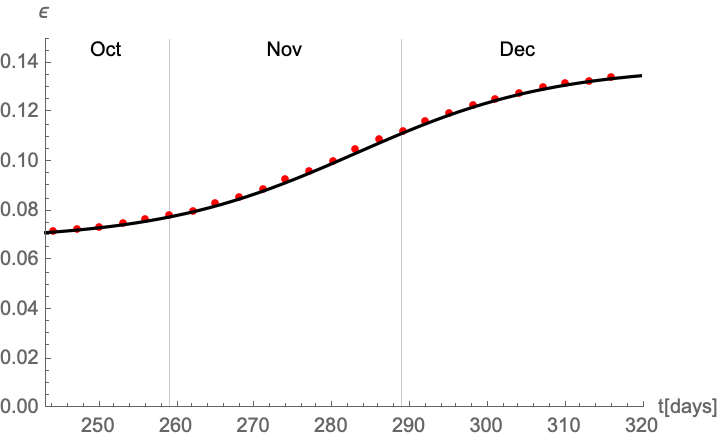}\\
\includegraphics[width=7.5cm]{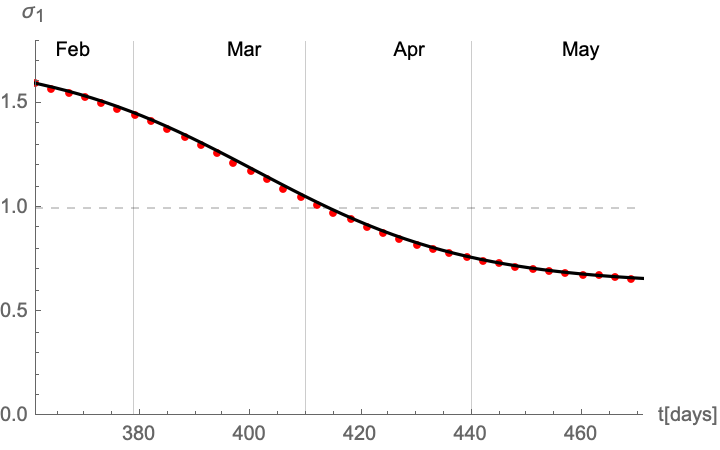}\hspace{1cm}\includegraphics[width=7.5cm]{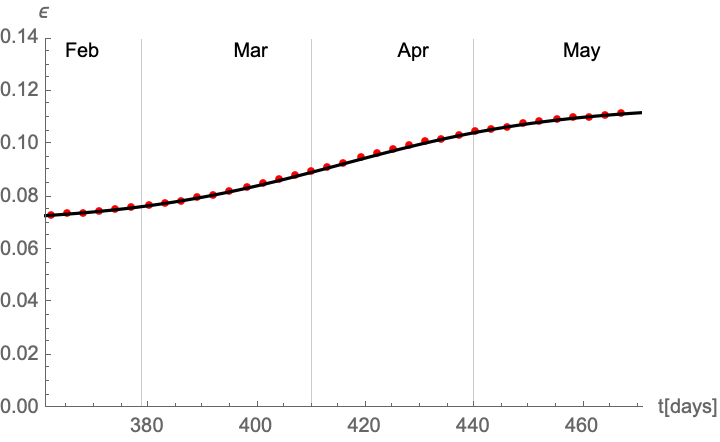}\\
\end{center}
\caption{The time-dependent parameters $(\sig_1,\rhe)$ for the compartmental model (\ref{DiffSIIRV}) (red dots) along with their interpolations (black lines) following (\ref{ApproxTimeDepsGen}) needed to match the epidemiological data for the second (top panels) and third (bottom panels) waves in Austria. }
\label{Fig:ComparisonFitRealActivesAustria}
\end{figure}

The time-dependent functions $(\sig_1(t),\rhe(t))$ that are needed to fit the data of waves 2 and 3 (where no \ep\, model was imposed) are shown in Figure~\ref{Fig:ComparisonFitRealActivesAustria}.

\subsubsection{Wave 4}

\begin{wrapfigure}{r}{0.45\textwidth}
\begin{center}
${}$\\[-1cm]
\includegraphics[width=7.5cm]{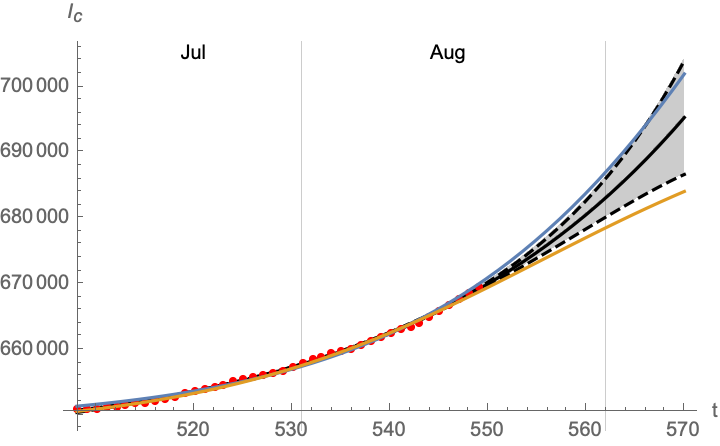}
\end{center}
\caption{\footnotesize Cumulative number of infected in Austria starting from July/2021.}
\label{Fig:FittingWave4Austria}
${}$\\[-3.5cm]
\end{wrapfigure}

The infection numbers starting from 07/July are shown in Figure~\ref{Fig:FittingWave4Austria} along with a 90\% confidence 

\begin{figure}[htbp]
\begin{center}
\includegraphics[width=7.5cm]{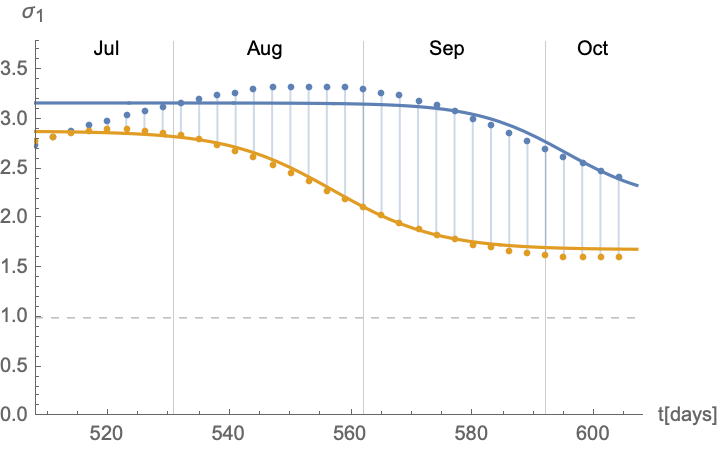}\hspace{1cm}\includegraphics[width=7.5cm]{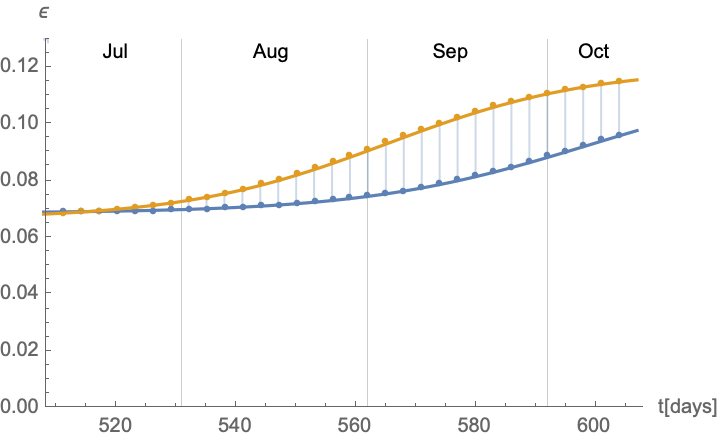}\\
\includegraphics[width=7.5cm]{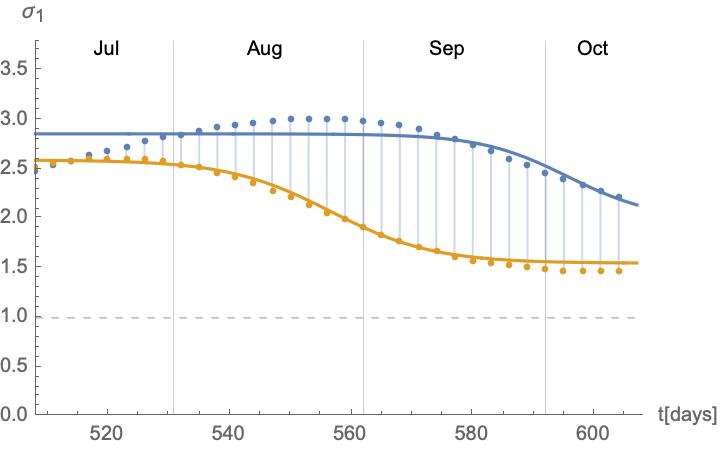}\hspace{1cm}\includegraphics[width=7.5cm]{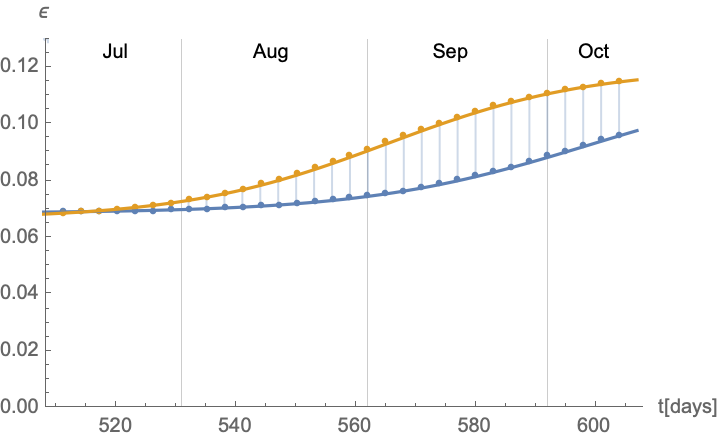}\\
\end{center}
\caption{Time dependent parameters $\sig_1$ (left panel) and $\rhe$ (right panel) for the extremal cases of wave 4. The blue and orange colours are correlated with the curves in the right panel of Figure~\ref{Fig:FittingWavesGermany}. The panels of the top row use $p=0.8$, while the panels in the bottom row use $p=0.9$ in order to fit the data.}
\label{Fig:ComparisonFitWave4Aus}
\end{figure}

\noindent 
interval. Furthermore the coloured curves are approximate fits with a logistic function with the following parameters
\begin{align}
&A_0^+=200000\,,&&\lam_0^+=0.053\,,\nonumber\\
&t_0^+=589.1\,,&&\delta_0^+=648808\,,\nonumber\\
&A_0^-=51856\,,&&\lam_0^-=0.059\,,\nonumber\\
&t_0^-=555.3\,,&&\delta_0^-=647643\,.\label{AustriaExtremalCases}
\end{align} 
These data suggest the onset of a new wave, just as in the case of Germany. However, unlike Germany, a \vtep\, (3-G-rule: 'geimpft, getestet, genesen') was enforced on 01/July/2021 (with earlier measures dating as far back as 19/May/2021) \cite{AustMinisterium} \footnote{The rule was slightly modified on 22/07 and 15/08 specifying stricter rules to discotheques and nightclubs and imposing restrictions to only partially vaccinated individuals respectively. See Appendix~\ref{App:ExamplesPass} for further details.} allowing individuals full access to the public life only with a certificate of either being (fully) vaccinated, having recovered from a previous infection or having tested negative for \cov\,. Therefore, in order to derive time-dependent parameters $(\sig_1(t),\rhe(t))$ for the extremal cases (\ref{AustriaExtremalCases}), we need to take the presence of the \vep\, into account. Since the exact efficacy of the Green Pass are difficult to quantify, we have used $p=0.8$ and $p=0.9$ as reference values to fit the data. The corresponding time-dependent functions $(\sig_1(t),\rhe(t))$ for wave 4 are shown in Figure (\ref{Fig:ComparisonFitWave4Aus}). As is evident, the main difference lies in the function $\sig_1$, while the curve for $\rhe$ is relatively unchanged.\footnote{As remarked before, mathematically the parameter $\pvt$ can be absorbed in the $\rin_1$. Since we assumed for the latter anyway a certain range, the main effect of $\pvt$ can also be absorbed in the quotient $\sig_2/\sig_1$, \emph{i.e.} the reduction in the rate at which vaccinated infectious individuals infect others.}

\begin{figure}[htbp]
\begin{center}
\includegraphics[width=7.5cm]{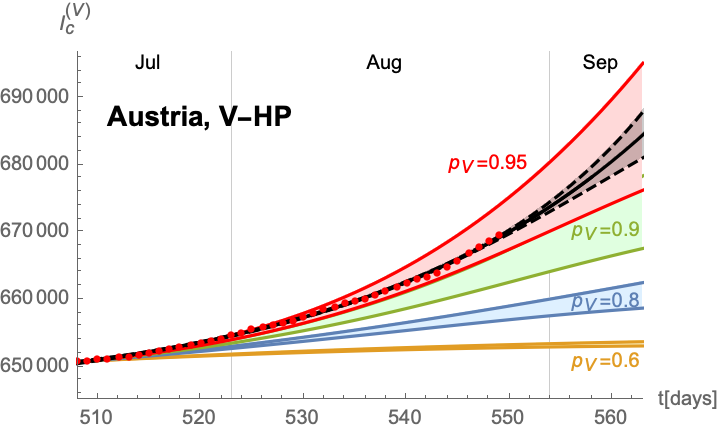}\hspace{1cm}\includegraphics[width=7.5cm]{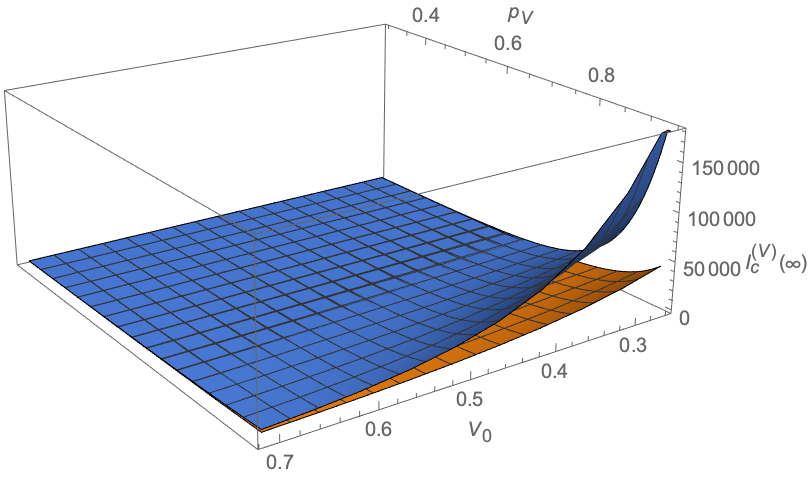}
\end{center}
\caption{Predicted time evolution of the cumulative number of infected individuals for different values of the efficacy $\pv$ of the \vep\,model. Left panel: time evolution, right panel: asymptotic cumulative number of infected individuals as a function of $p_V$ and $V_0$. Both plots assume $\sig_2/\sig_1=1$, $\rvac=0.0085$ as well as $\pvt=0.9$ for the \vtep\, currently in place in Austria.}
\label{Fig:GreenPassAustria09}
\end{figure}

\begin{figure}[htbp]
\begin{center}
\includegraphics[width=7.5cm]{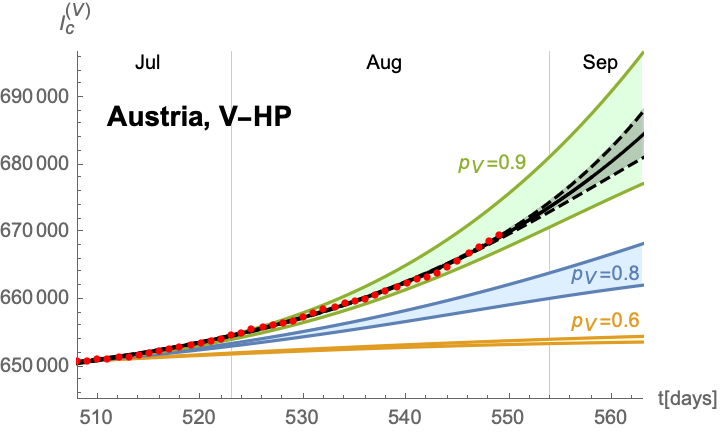}\hspace{1cm}\includegraphics[width=7.5cm]{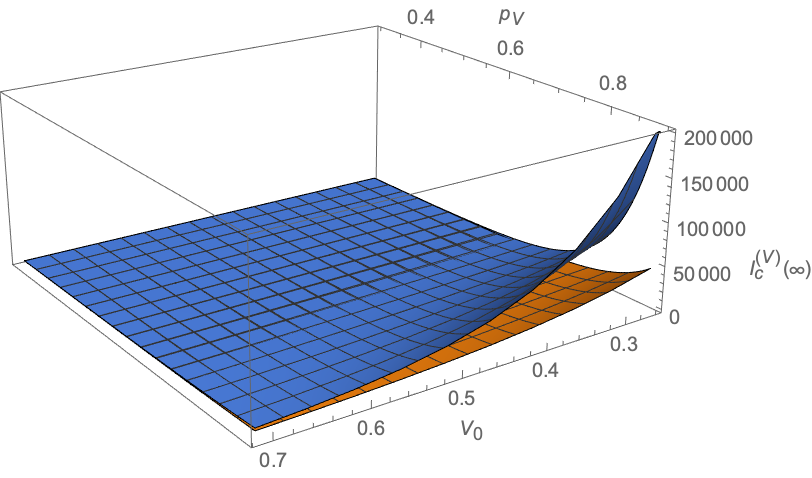}
\end{center}
\caption{Predicted time evolution of the cumulative number of infected individuals for different values of the efficacy $\pv$ of the \vep\,model. Left panel: time evolution, right panel: asymptotic cumulative number of infected individuals as a function of $p_V$ and $V_0$. Both plots assume $\sig_2/\sig_1=1$, $\rvac=0.0085$ as well as $\pvt=0.8$ for the \vtep\, currently in place in Austria.}
\label{Fig:GreenPassAustria08}
\end{figure}

\noindent
The final step consists in applying the time-dependent $(\sig_1,\rhe)$ parameters to the stronger \vep\, model (\ref{DiffSIIRVgreenpass}). The results are shown in Figure~\ref{Fig:GreenPassAustria09} for $\pvt=0.9$ and Figure~\ref{Fig:GreenPassAustria08} for $\pvt=0.8$ respectively. The results again support an equivalence of the type (\ref{EquivRelPvPvt}) between the parameters $\pv$ of the \vep\, and $\pvt$ of the \vtep\,model.

\subsection{Further Examples}
In the following we briefly show the results of a similar analysis as in the previous section for France, Denmark and Italy, which have similar \vtep\, models in place as Austria. For more details on the concrete implementations of the \vtep\,, see Appendix~\ref{App:ExamplesPass}.
\subsubsection{France}
As a further example we consider France, where since 09/08 a 'pass sanitaire' is obligatory, which follows roughly the \vtep\, model (\ref{DiffSIIRVgreenpass3G}). Since thus the situation is to some extent similar as in Austria, we focus our discussion on a potential fourth wave in the end of summer/beginning of fall. At the time when this paper was being finalised, the actual infection numbers in France have been dropping, indicating the end of this wave. While this assessment neglects possible new developments related to the opening of schools and universities in September, we can fit 

\begin{wrapfigure}{r}{0.5\textwidth}
\begin{center}
${}$\\[-0.5cm]
\includegraphics[width=7.5cm]{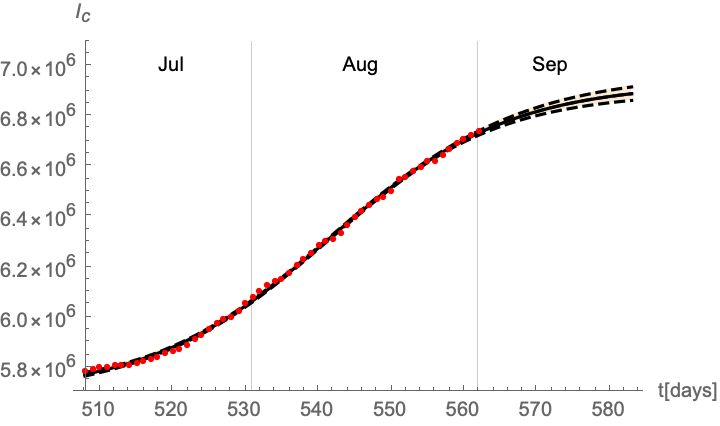}
\caption{\footnotesize Cumulative number of infected in France}
\label{Fig:FranceWave4}
\end{center}
${}$\\[-1.5cm]
\end{wrapfigure}

\noindent
the data under this assumption rather precisely with a logistic function, as is shown in Figure~\ref{Fig:FranceWave4}, with only a narrow 90\% confidence interval.

\begin{figure}[htbp]
\begin{center}
\includegraphics[width=7.5cm]{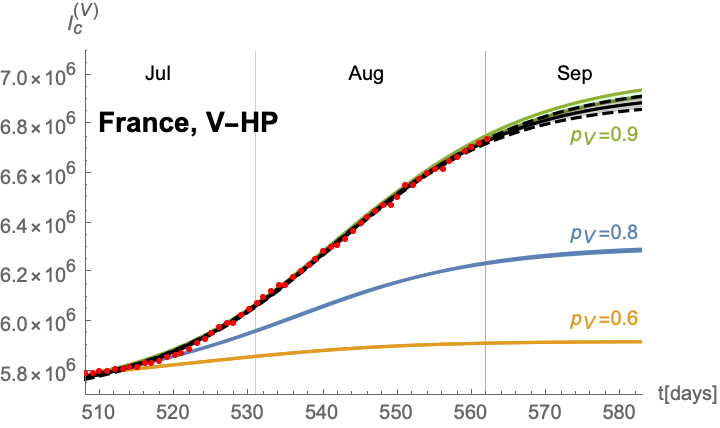}\hspace{1cm}\includegraphics[width=7.5cm]{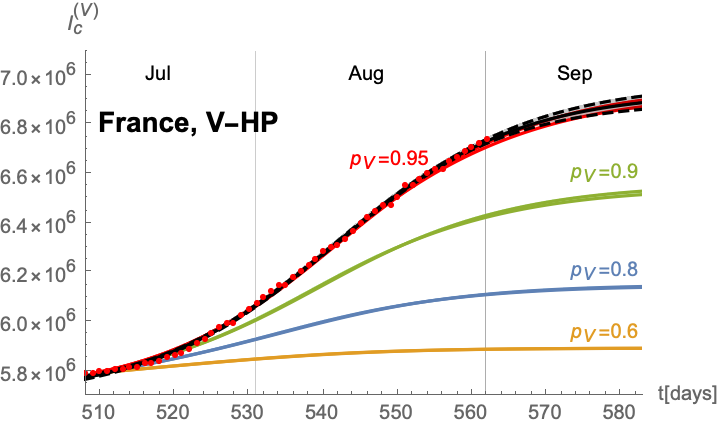}
\end{center}
\caption{Predicted time evolution of the cumulative number of infected individuals for different values of the efficacy $\pv$ of the \vep\,model. Both plots assume $\sig_2/\sig_1=1$, as well as $\pvt=0.8$ (left panel) and $\pvt=0.9$ (right panel) for the \vtep\, currently in place in Austria.}
\label{Fig:GreenPassFrance}
\end{figure}

We again consider the two cases that the current pass-sanitaire has an efficiency of $\pvt=0.8$ and $\pvt=0.9$ and compare the results with the stronger \vep. Indeed repeating the same analysis as in the previous subsection yields the plots in Figures~\ref{Fig:GreenPassFrance}. The results again support an equivalence of the type (\ref{EquivRelPvPvt}) between the parameters $\pv$ of the \vep\, and $\pvt$ of the \vtep\,model.

\subsubsection{Denmark}
As a further example we consider Denmark, where  a \vtep\, (called Corona-Passport) of the  

\begin{wrapfigure}{r}{0.5\textwidth}
\begin{center}
${}$\\[-0.5cm]
\includegraphics[width=7.5cm]{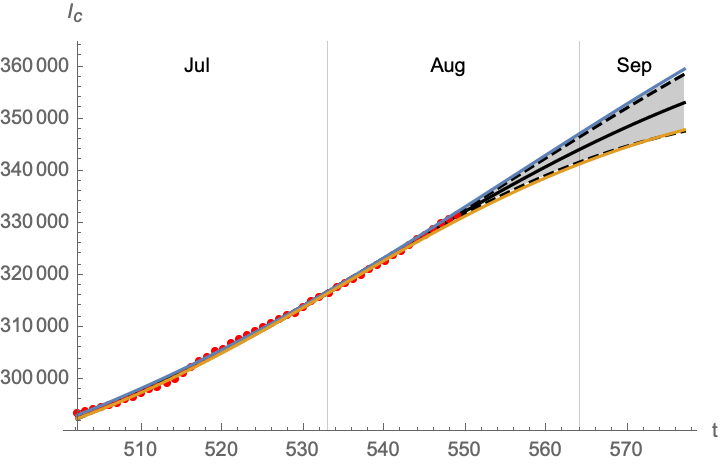}
\caption{\footnotesize Cumulative number of infected in Denmark}
\label{Fig:DenmarkWave4}
\end{center}
${}$\\[-1.5cm]
\end{wrapfigure}

\noindent
form~(\ref{DiffSIIRVgreenpass3G}) has been imposed, which follows roughly the model (\ref{DiffSIIRVgreenpass3G}).  As in the case of France, we focus exclusively on the development in the summer 2021: the epidemiological data along of a fit with a logistic function are shown in Figure~\ref{Fig:DenmarkWave4}. 

\begin{figure}[htbp]
\begin{center}
\includegraphics[width=7.5cm]{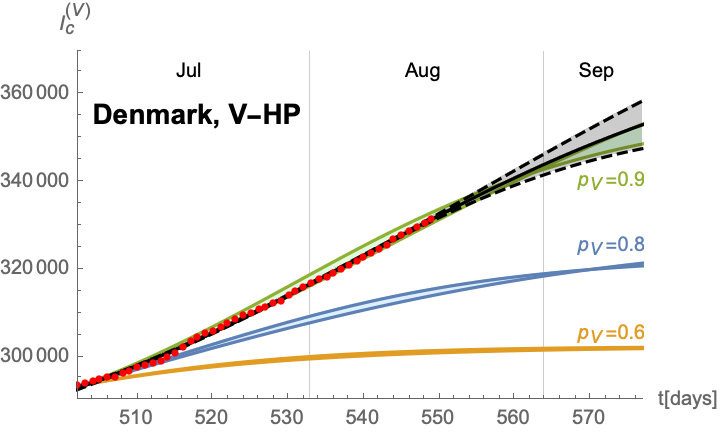}\hspace{1cm}\includegraphics[width=7.5cm]{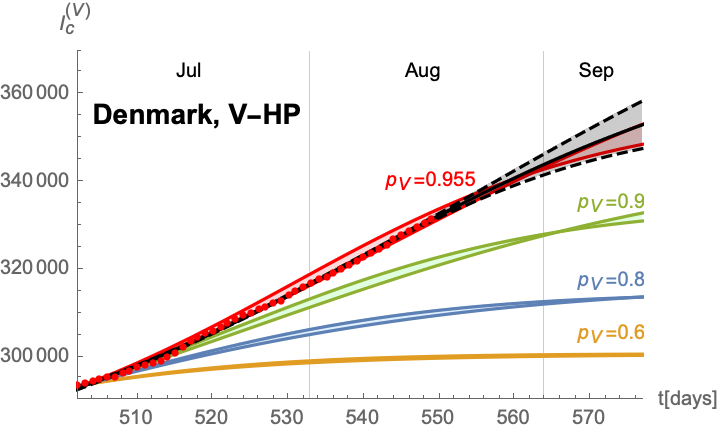}
\end{center}
\caption{Predicted time evolution of the cumulative number of infected individuals for different values of the efficacy $\pv$ of the \vep\,model. Both plots assume $\sig_2/\sig_1=1$, as well as $\pvt=0.8$ (left panel) and $\pvt=0.9$ (right panel) for the \vtep\, in Denmark.}
\label{Fig:GreenPassDenmark}
\end{figure}

In absence of studies estimating the efficiency of the 'Corona Passport', we consider the two cases of $\pvt=0.8$ and $\pvt=0.9$ and compare the results with a stronger \vep. Indeed repeating the same analysis as in the previous subsection yields the plots in Figures~\ref{Fig:GreenPassDenmark}. The results again support an equivalence of the type (\ref{EquivRelPvPvt}) between the parameters $\pv$ of the \vep\, and $\pvt$ of the \vtep\,model.

\subsubsection{Italy}
As a final example we consider Italy, where  a \vtep\, of the form~(\ref{DiffSIIRVgreenpass3G}), has been imposed.

\begin{wrapfigure}{r}{0.5\textwidth}
\begin{center}
${}$\\[-0.5cm]
\includegraphics[width=7.5cm]{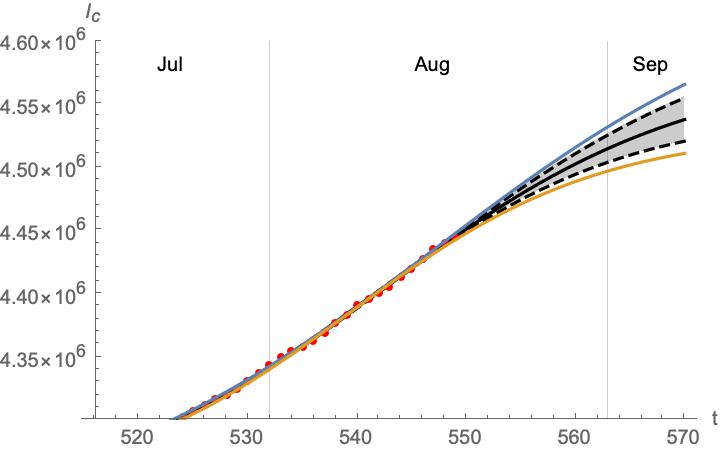}
\caption{\footnotesize Cumulative number of infected in Italy}
\label{Fig:ItalyWave4}
\end{center}
${}$\\[-2cm]
\end{wrapfigure}

\noindent
 As in the previous cases, we focus exclusively on the development in the summer 2021: the epidemiological data along of a fit with a logistic function are shown in Figure~\ref{Fig:ItalyWave4}. 

\begin{figure}[htbp]
\begin{center}
\includegraphics[width=7.5cm]{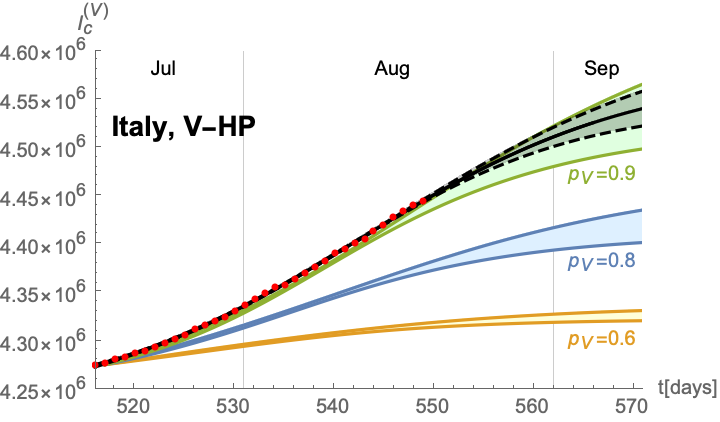}\hspace{1cm}\includegraphics[width=7.5cm]{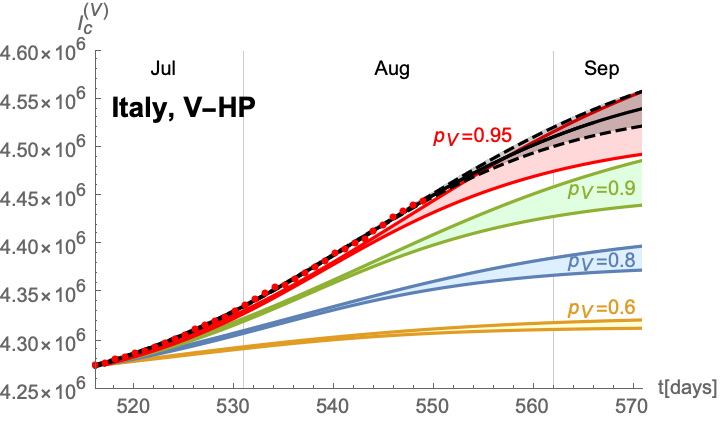}
\end{center}
\caption{Predicted time evolution of the cumulative number of infected individuals for different values of the efficacy $\pv$ of the \vep\,model. Both plots assume $\sig_2/\sig_1=1$, as well as $\pvt=0.8$ (left panel) and $\pvt=0.9$ (right panel) for the \vtep\, currently in place in Ital.}
\label{Fig:GreenPassItaly}
\end{figure}

In absence of studies estimating the efficacy of the \vtep\,currently in place in Italy, we consider the two cases of $\pvt=0.8$ and $\pvt=0.9$ and compare the results with a stronger \vep. Indeed repeating the same analysis as in the previous subsection yields the plots in Figures~\ref{Fig:GreenPassItaly}. The results again support an equivalence of the type (\ref{EquivRelPvPvt}) between the parameters $\pv$ of the \vep\, and $\pvt$ of the \vtep\,model.

\section{Conclusions}\label{Sect:Conclusions}
In this paper we have analysed the impact of so called Health Passes on the epidemiological dynamics of infectious diseases. These \ep s correspond to measures that restrict the access of individuals with a higher risk of being infectious to public life. Concretely, we have distinguished two different classes that grant access to individuals with a vaccination certificate or a recent negative test (\vtep) and only to vaccinated individuals (\vep). 

We have first discussed these \ep s in the context of a simple compartmental SIIRV model (\ref{DiffSIIRV}) and have generalised them in the context of the eRG framework, which is better suited for describing the dynamics over a longer period of time, in particular an entire epidemiological wave. Indeed, analysing in particular the dependence of the asymptotic cumulative number of infected individuals (which is a crucial parameter in the description of the eRG), we have found the approximative exponential dependence (\ref{SolpImp}) on the parameter describing the efficacy of the \ep\,. Furthermore, comparing the efficacy of a \vtep-model to a \vep\, model reduces to comparing the corresponding $\theta$-parameters appearing in this approximation.

We have furthermore validated our models by discussing the diffusion of COVID-19 in several European countries. We have analysed in detail Germany (who, to this date, has not implemented any \ep) and Austria (who currently has implemented a \vtep\, and considers the partial introduction of a \vep) and have presented a briefer analysis for France, Denmark and Italy. In all cases we have established that a \vep\, is much more efficient in reducing the number of infected. Our model in fact allows for a quantitative comparison, leading to the relation (\ref{EquivRelPvPvt}): if all remaining parameters remain the same, the efficacy of a \vtep\, needs to roughly be twice as high to produce the same reduction of infections as a \vep\,. Furthermore  in most cases, an efficiency of a \vep\, of roughly 20-40\% is strong enough to completely suppress a potential fourth wave. 

We have undertaken preliminary studies that also include a potential reduction in the number of tests (related to a reduction in the removal rate due to a reduced capacity of identifying and isolating infected individuals). It would be important to further extend these studies, in particular to establish a quantitative relation between these two rates.

\newpage 
\appendix

\section{SIIRV Model for Vanishing Vaccination Rate}
\subsection{Decoupling the SIIRV Model}
While the differential equations (\ref{DiffSIIRV}) are in general difficult to solve analytically, there are a few results we can derive in the case of vanishing vaccination rate: we start with the system (\ref{DiffSIIRV}) for $\rvac=0$ and divide the differential equation for $\frac{d\ms}{dt}$ by the differential equation for $\frac{d\mv}{dt}$
\begin{align}
\frac{d\ms}{d\mv}=\frac{\ms}{\riv\mv}\,,
\end{align}
which has the following solution compatible with the initial conditions
\begin{align}
\frac{\mv(t)}{\mv_0}=\left(\frac{\ms(t)}{\ms_0}\right)^\riv\,,&&\forall\, t\in\mathbb{R}\,.\label{RelSV}
\end{align}
Inserting this relation into the differential equations for $\frac{d\mi_{1}}{dt}$ and $\frac{d\mi_{2}}{dt}$ in (\ref{DiffSIIRV}) we obtain
\begin{align}
&\frac{d\mi_1}{dt}=\ms\left[\rin_1\,\mi_1+\rin_2\,\mi_2\right]-\rhe\,\mi_1\,,&&\text{and} &&\frac{d\mi_2}{dt}=\riv\,\mv_0\,\left(\frac{\ms}{\ms_0}\right)^\riv\,\left[\rin_1\,\mi_1+\rin_2\,\mi_2\right]-\rhe\,\mi_2\,.\label{SeparatedI}
\end{align}
The combination of these two equations implies
\begin{align}
\frac{d(\rin_1\mi_1+\rin_2\mi_2)}{dt}=\left(\rin_1\,\ms+\rin_2\riv\,\mv_0\,\left(\frac{\ms}{\ms_0}\right)^\riv-\rhe\right)\,(\rin_1\mi_1+\rin_2\mi_2)\,,\label{DiffEqI1I2}
\end{align}
which has the following solution compatible with the initial conditions
\begin{align}
(\rin_1\mi_1+\rin_2\mi_2)(t)=(\rin_1\mi_{1,0}+\rin_2\mi_{2,0})\,\text{exp}\left[\int_0^t dt'\,\left(\rin_1\,\ms(t')+\rin_2\,\riv\,\mv_0\left(\frac{\ms(t')}{\ms_0}\right)^\riv-\rhe\right)\right]\,.\label{CombineIs}
\end{align}
Inserting this result into the first equation of (\ref{DiffSIIRV}) gives an integro-differential equation for $\ms$
\begin{align}
\frac{d\ms}{dt}(t)=-\ms(t)\,(\rin_1\mi_{1,0}+\rin_2\mi_{2,0})\,\text{exp}\left[\int_0^t dt'\,\left(\rin_1\,\ms(t')+\rin_2\,\riv\,\mv_0\left(\frac{\ms(t')}{\ms_0}\right)^\riv-\rhe\right)\right]
\end{align}
which leads to the following (non-linear) second order differential equation 
\begin{align}
\frac{d}{dt}\left(\ln\left(\frac{d\ln S}{dt}\right)\right)=\rin_1\,\ms+\rin_2\,\riv\,\mv_0\left(\frac{\ms}{\ms_0}\right)^\riv-\rhe\,.\label{DiffEqSecondS}
\end{align}
The latter yields a solution for $\ms$, which, when injected together with (\ref{CombineIs}) into the first equation of (\ref{SeparatedI}) yields an ordinary differential equation for $\mi_1$, thus decoupling the initial system (\ref{DiffSIIRV}).

\subsection{Herd Immunity}\label{App:HerdImmunitySIIRV}
From the results of the previous subsection, we can derive an analytical expression for the herd immunity threshold discussed in Section~\ref{Sect:HerdSIIRV}. To this end, we first derive the condition for the (relative) number of new infected to reach a local extremum. Indeed, the former can be defined as the time derivative of $\Ic(t)$
\begin{align}
\Icnew(t)=\frac{1}{N}\,\frac{d\Ic}{dt}(t)=\frac{d}{dt}(1-\ms-\mv)\,.
\end{align}
A necessary condition for a local extremum of the (relative) number of new cases is therefore
\begin{align}
0&=\frac{d^2}{dt^2}(\ms+\mv)=\frac{d^2\ms}{dt^2}+\mv_0\,\frac{d^2}{dt^2}\left(\frac{\ms}{\ms_0}\right)^\riv=\frac{d^2\ms}{dt^2}+\frac{\mv_0\riv}{\ms_0}\left[\frac{\ms}{\ms_0}\,\frac{d^2\ms}{dt^2}+\frac{\riv-1}{\ms_0}\left(\frac{d\ms}{dt}\right)^2\right]\left(\frac{\ms}{\ms_0}\right)^{\riv-2}\,,\nonumber\\
&=\left(1+\riv\frac{\mv}{\ms}\right)\,\frac{d^2\ms}{dt^2}+\riv(\riv-1)\frac{\mv}{\ms^2}\,\left(\frac{d\ms}{dt}\right)^2\,.
\end{align}
where we have used (\ref{RelSV}). We next use the differential equation (\ref{DiffEqSecondS}) to eliminate $\frac{d^2\ms}{dt^2}$
\begin{align}
0&=\left(1+\riv\frac{\mv}{\ms}\right)\,\left[\frac{1}{\ms}\left(\frac{d\ms}{dt}\right)+\rin_1\,\ms+\rin_2\,\riv\,\mv-\rhe\right]\left(\frac{d\ms}{dt}\right)+\riv(\riv-1)\frac{\mv}{\ms^2}\,\left(\frac{d\ms}{dt}\right)^2\nonumber\\
&=\frac{1}{S}\left(\frac{d\ms}{dt}\right)\left[\ms\left(\rin_1\,\ms+\rin_2\,\riv\,\mv-\rhe+\frac{1}{\ms}\frac{d\ms}{dt}\right)+\riv\mv\left(\rin_1\,\ms+\rin_2\,\riv\,\mv-\rhe+\frac{\riv}{\ms}\frac{d\ms}{dt}\right)\right]
\end{align}
Since we are looking for a local extremum for $t<\infty$, we may assume $\ms(t)> 0$ and $\mv(t)>0$. Therefore, the previous condition has two solutions
\begin{itemize}
\item $\frac{d\ms}{dt}=0$: with (\ref{RelSV}) this condition also implies $\frac{d\mv}{dt}=0$ and thus with (\ref{DiffSIIRV})
\begin{align}
&\ms(t)=0=\mv(t)\,,&&\text{or} &&\mi_1(t)=0=\mi_2(t)\,,
\end{align}
(since $\mi_{i}\geq 0$ for $i=1,2$). The former relation corresponds to the case where the entire population has been infected, while the latter case corresponds to the eradication of the disease. Both cases constitute the end of the epidemic and are thus not local extrema
\item the relation
\begin{align}
0=\ms\left(\rin_1\,\ms+\rin_2\,\riv\,\mv-\rhe+\frac{1}{\ms}\frac{d\ms}{dt}\right)+\riv\mv\left(\rin_1\,\ms+\rin_2\,\riv\,\mv-\rhe+\frac{\riv}{\ms}\frac{d\ms}{dt}\right)\label{CondComp}
\end{align}
Since we asssume $\ms>0$ and $\mv>0$, this relation can only hold if exactly one of the terms
\begin{align}
&\rin_1\,\ms+\rin_2\,\riv\,\mv-\rhe+\frac{1}{\ms}\frac{d\ms}{dt}\,,&&\text{and} &&\rin_1\,\ms+\rin_2\,\riv\,\mv-\rhe+\frac{\riv}{\ms}\frac{d\ms}{dt}\,,
\end{align}
is negative (and the other positive). However, since $\mi_{1,2}>0$ and therefore (\ref{DiffSIIRV}) implies that $\ms$ is a monotonically decreasing function, \emph{i.e.} $\frac{d\ms}{dt}<0$, this is only possible if 
\begin{align}
\rin_1\,\ms+\rin_2\,\riv\,\mv-\rhe>0\,,
\end{align}
which leads to the following necessary condition\footnote{For $\riv=0$, equation (\ref{CondComp}) becomes $\ms(\rin_1\,\ms-\rhe)=0$ leading to the necessary condition $\rin_1\,\ms_0\geq \rhe$ for a local extremum, which is compatible with (\ref{CondNecessary}).}
\begin{align}
\rin_1\,\ms_0+\rin_2\,\riv\,\mv_0>\rhe\,.\label{CondNecessary}
\end{align}
\end{itemize}
The relation (\ref{CondNecessary}) can also be formulated in terms of (\ref{DefSigs})
\begin{align}
\sig_1\,\ms_0+\riv\,\sig_2\,\mv_0>1\,.\label{BoundCond}
\end{align}
Notice, for $\zeta=0$ (in which case $\mv$ is decoupled from the time evolution), this condition is compatible with the usual threshold condition $\sig_1\,\ms_0>1$ of the SIR model.

In order to define the herd immunity threshold (HIT), we first define the vaccinated fraction of the population at the outbreak of the disease as
\begin{align}
\herd=\frac{\mv_0}{\ms_0+\mv_0}\,.
\end{align}
We furthermore consider initial conditions corresponding to the limit $\mi_{1,0}\to0$, such that $1=\ms_0+\mv_0$ and the threshold $\hit=\mv_0$ such that the number of new infected individuals does not reach an extremum (but is monotonically decreasing). From (\ref{BoundCond}) we then obtain the limiting case
\begin{align}
\sig_1\,(1-\hit)+\riv\,\sig_2\,\hit=1\,,
\end{align}
leading to 
\begin{align}
\hit=\frac{\sig_1-1}{\sig_1-\riv\sig_2}\,,\label{HIT}
\end{align}
which in the limit $\riv\to 0$ (\emph{i.e.} a perfect vaccine), reduces to the usual definition (see \emph{e.g.} \cite{HIT})
\begin{align}
\lim_{\zeta\to 0}\hit=1-\frac{1}{\sig_1}\,.
\end{align}

\subsection{Analytic Asymptotics}
In order to get information about the asymptotic behaviour of the number of susceptible, we return to eq.~(\ref{DiffEqI1I2}) and divide by the first equation of (\ref{DiffSIIRV})
\begin{align}
\frac{d(\rin_1\,\mi_1+\rin_2\,\mi_2)}{dS}=-\frac{1}{S}\,\left(\rin_1\,\ms+\rin_2\riv\,\mv_0\,\left(\frac{\ms}{\ms_0}\right)^\riv-\rhe\right)\,.
\end{align}
Instead of a differential equation for the combination $\rin_1\,\mi_1+\rin_2\,\mi_2$ as a function of time, we have an equation as a function of $\ms$, which has the following solution that is compatible with the initial conditions 
\begin{align}
(\rin_1\,\mi_1+\rin_2\,\mi_2)=\rin_1\,\mi_{1,0}+\rin_2\,\mi_{2,0}-\rin_1\,(\ms-\ms_0)-\rin_2\,(\mv-\mv_0)+\rhe\,\ln\left(\frac{\ms}{\ms_0}\right)\,.
\end{align}
For $t\to \infty$ we have 
\begin{align}
\lim_{t\to\infty}\mi_{1}(t)=0=\lim_{t\to\infty}\mi_{2}(t)
\end{align}
such that we obtain the following equation for $\ms_\infty=\lim_{t\to \infty}\ms(t)$
\begin{align}
\rin_1\,\mi_{1,0}+\rin_2\,\mi_{2,0}=\rin_1\,(\ms_\infty-\ms_0)+\rin_2\,\mv_0\left[\left(\frac{\ms_\infty}{\ms_0}\right)^\zeta-1\right]-\rhe\,\ln\left(\frac{\ms_\infty}{\ms_0}\right)\,,
\end{align}
which can only be solved numerically for $\ms_\infty$.
\section{Asymptotic Expansion for the \ep-SIR Model}\label{App:ComparSIR}
To get an intuition about the asymptotic behaviour of the \ep-models (\ref{DiffSIIRVgreenpass3G}) and (\ref{DiffSIIRVgreenpass}), we consider as a simpler model the following two SIR models with a \ep  
\begin{align}
&\frac{d\ms}{dt}=-\pvt\,\rin\,\ms\,\mi\,,&\frac{d\mi}{dt}=\pvt\,\rin\,\ms\,\mi-\rhe\,\mi\,,&&\frac{d\mr}{dt}=\rhe\,\mi\,,&&\text{with}&&\begin{array}{l}\ms(t=0)=\ms_0\,,\\ \mi(t=0)=\mi_0\,,\\\mr(t=0)=0\,,\end{array}\label{SIR3G}
\end{align}
and 
\begin{align}
&\frac{d\ms}{dt}=-\pv^2\,\rin\,\ms\,\mi\,,&\frac{d\mi}{dt}=\pv^2\,\rin\,\ms\,\mi-\rhe\,\mi\,,&&\frac{d\mr}{dt}=\rhe\,\mi\,,&&\text{with}&&\begin{array}{l}\ms(t=0)=\ms_0\,,\\ \mi(t=0)=\mi_0\,,\\\mr(t=0)=0\,,\end{array}\label{SIR}
\end{align}
with constant $(\rin,\rhe)$. These correspond to the usual SIR model (in the absence of vaccinations) in which the infection rate has been rescaled by some power of a parameter $\pv,\pvt\in[0,1]$. 

\begin{wrapfigure}{r}{0.5\textwidth}
\begin{center}
${}$\\[-0.5cm]
\includegraphics[width=7.5cm]{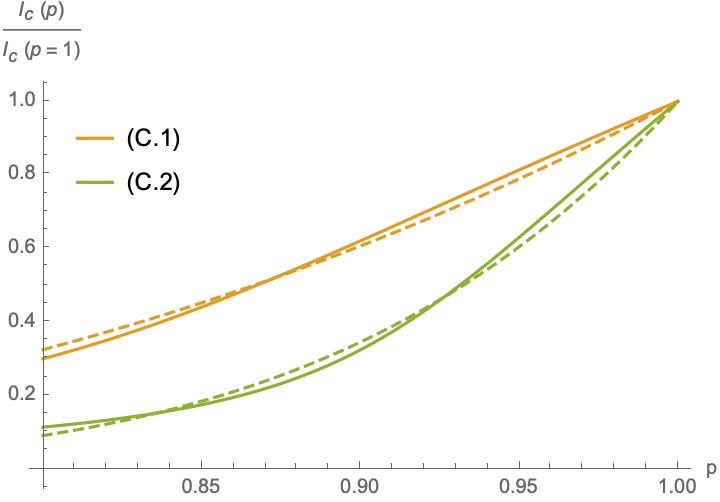}
\caption{\footnotesize Asymptotic cumulative number of infected individuals for the \ep-SIR models (\ref{SIR3G}) and (\ref{SIR}) as a function of $p_{\text{VT},\text{V}}$ (indicated as $p$) and normalised to $\pvt=1=\pv$. The solid orange curve is the asymptotic solution of the model (\ref{SIR3G}) and the solid green curve the solution of the model (\ref{SIR}). The dashed lines are approximations of the solutions according to (\ref{ExpoApprox}) (with $\theta=9.56$ for (\ref{SIR3G}) and $\theta=4.49$ for (\ref{SIR})). The plots use $\ms_0=0.99$ and $\sig=1.2$.}
\label{Fig:gSIRPlots}
\end{center}
${}$\\[-2cm]
\end{wrapfigure}

\noindent
Although mathematically this is a rather trivial rescaling, we can learn certain qualitative features of the different power of the parameter $p$, which we shall also observe in the more complicated models in the main text of this paper that also accommodate vaccinations. Indeed, in absence of a vaccination dynamics, the asymptotic cumulative number of infected for the systems (\ref{SIR}) can be computed analytically
\begin{align}
&\frac{\Icml(\infty,\pvt)}{N}=1+\frac{W\left(-\ms_0\,\sig\,\pvt\,e^{-\sig \pvt}\right)}{\sig \pvt}\,,\nonumber\\
&\frac{\Icms(\infty,\pv)}{N}=1+\frac{W\left(-\ms_0\,\sig\,\pv^2\,e^{-\sig \pv^2}\right)}{\sig \pv^2}\,,\label{AsymSIR}
\end{align}
with $\sig=\frac{\rin}{\rhe}$ and where $W$ is the Lambert function. $\Ic^{(\text{VT},\text{V})}(\infty,p)$ as a function of $p_{\text{VT},\text{V}}$ (and normalised to $p=1$) is shown in Figure~\ref{Fig:gSIRPlots}.

\begin{figure}[htbp]
\begin{center}
\includegraphics[width=7.5cm]{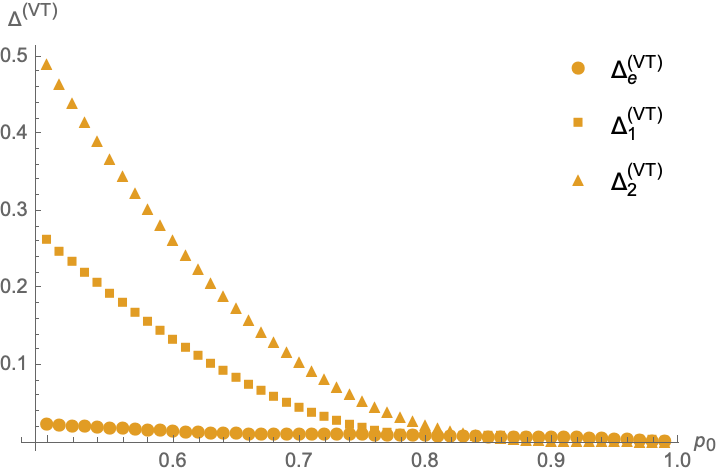}\hspace{1cm}\includegraphics[width=7.5cm]{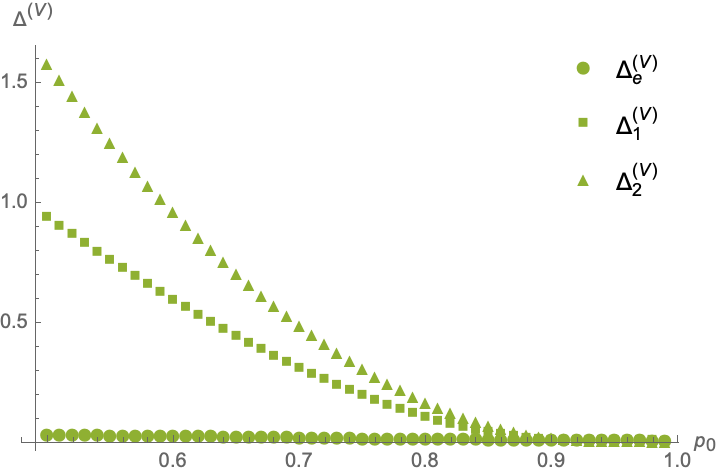}
\end{center}
\caption{Comparison of the quadratic error (\ref{DefQuadError}) for the approximation (\ref{ExpoApprox}) ($\Delta_e^{(\text{VT},\text{V})}$, indicated by the circles) with the error for the first order ($\Delta_1^{(\text{VT},\text{V})}$, indicated by the squares) and second order ($\Delta_1^{(\text{VT},\text{V})}$, indicated by the triangles) Taylor series expansion around $p=1$ as a function of $p_0$. The left panel shows the estimation for the model (\ref{SIR3G}) and the right panel for (\ref{SIR}). Both plots use $S_0=0.99$ and $\sig=1.2$.}
\label{Fig:ErrorSIR}
\end{figure}

As can be seen from Figure~\ref{Fig:gSIRPlots}, for values of $p_{1,3}$ close to 1, the normalised asymptotic number of infected can be approximated by an exponential function of the form
\begin{align}
&\frac{\Ic(\infty,p)}{\Ic(\infty,p=1)}\sim\text{exp}\left(\theta\,\frac{p-1}{p}\right)\,,&&\text{with} &&\theta\in\mathbb{R}\,,\label{ExpoApprox}
\end{align}
for a constant $\theta$ that depends on $\ms_0$ and $\sigma$. In Figure~\ref{Fig:ErrorSIR} the quadratic error of the approximation~$\mathcal{E}$
\begin{align}
\Delta_{\mathcal{E}}^{(\text{VT},\text{V})}=\sqrt{\int_{p_0}^1dp\,\left[\frac{\Ic^{(\text{VT},\text{V})}(\infty,p)}{\Ic^{(\text{VT},\text{V})}(\infty,1)}-\mathcal{E}\right]^2}\,,\label{DefQuadError}
\end{align}
as a function of $p_0$ is compared for $\mathcal{E}$ given in (\ref{ExpoApprox}) with the first and second order of a Taylor series expansion around $p=1$. 

For the model (\ref{SIR}) a heuristic explanation of the approximation (\ref{ExpoApprox}) can be given as follows: let $A$ be the total number of infectious contacts throughout the entire time duration of the pandemic, such that for small initial conditions $A\sim \Icms(\infty)$. Considering these contacts for a value given value of $p$ close to one and changing it by a small $\delta p$, leads to a modification of $A$ that is (for small values of $p$) proportional to the number of infectious contacts for $p=1$, which is roughly $A(p)/p^2$. Thus we find that $A$ has to satisfy the approximate differential equation 
\begin{align}
\frac{dA}{dp}=\frac{\theta}{p^2}\,A(p)\,,
\end{align}
for some constant $\theta$, whose solution is indeed (\ref{ExpoApprox}). For the model (\ref{SIR3G}), the same argument

\begin{wrapfigure}{r}{0.5\textwidth}
\begin{center}
${}$\\[-0.5cm]
\includegraphics[width=7.5cm]{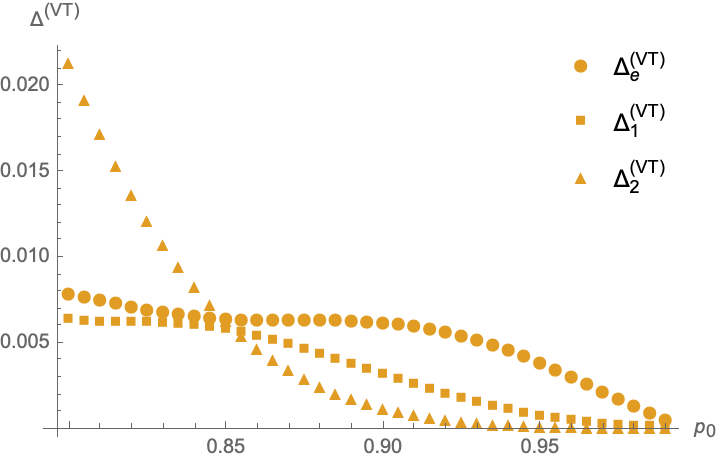}
\caption{\footnotesize Detailed estimation for $p_0$ closer to 1 of the model (\ref{SIR3G}). The parameters are as in the right plot of Figure~\ref{Fig:ErrorSIR}}
\label{Fig:SIR3GapproxDetail}
\end{center}
${}$\\[-2cm]
\end{wrapfigure}

\noindent
would lead to the modified differential equation 
\begin{align}
\frac{dA}{dp}=\frac{\theta}{p}\,A(p)\,,\label{DiffHeurist}
\end{align}
 which would suggest a linear behaviour. The plot in Figure~\ref{Fig:SIR3GapproxDetail} shows more details of the approximation for $p_0$ in the vicinity of 1, in which case the linear Taylor series is a better approximation than (\ref{ExpoApprox}). However, for $p_0$ further from $p=1$ the other dynamic of the system becomes more important and the heuristic argument leading to (\ref{DiffHeurist}) are no longer justified.
\section{SIIRV with Time-Dependent $\sig_1$ and Constant $\rhe$}\label{App:ConstEps}

\noindent
In this appendix we explore another approach to implementing time-dependent rates into the 

\begin{wrapfigure}{r}{0.5\textwidth}
\begin{center}
${}$\\[-0.5cm]
\includegraphics[width=7.5cm]{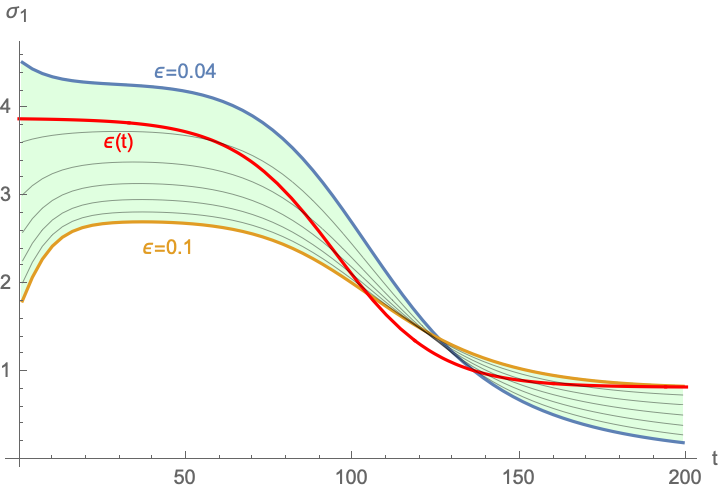}
\caption{\footnotesize Time dependence of the infection rate $\sig_1$ (left panel) for constant $\rhe\in[0.04,0.1]$ that is needed to reproduce $\Ic(t)$ of the form (\ref{LogisticFunction}) with the compartmental model (\ref{DiffSIIRV}). The red curve shows as comparison the $\sig_1$ obtained assuming a time-dependent $\rhe$ as in Figure~\ref{Fig:TimeDepRates}, using the same remaining parameters.}
\label{Fig:TimeDepSigConstEps}
\end{center}
${}$\\[-2cm]
\end{wrapfigure}

\noindent
SIIRV model (\ref{DiffSIIRV}) and compare their impact on the \vtep\, and \vep\, models (\ref{DiffSIIRVgreenpass3G}) and (\ref{DiffSIIRVgreenpass}) respectively. Indeed, in Section~\ref{Sect:TimeDepSIIRV} we have allowed both $(\sig_1,\rhe)$ to depend on time (leading to (\ref{ApproxTimeDepsGen})). This has required us to assume the number of infectious individuals (\ref{ApproxActiveI}) associated with (\ref{LogisticFunction}), which indeed correctly captured the numbers found in Germany and Austria (see Figures~\ref{Fig:FittingActivesGermany} and (\ref{Fig:FittingActivesAustria}) respectively). Similarly the time dependent $(\sig_1,\rhe)$ were similar to the results obtained in \cite{DellaMorte:2020qry} in the context of a simpler compartmental model. 

In this appendix, we shall rather make the assumption that $\rhe$ is a constant in time.This assumes that (at least at short times) the recovery rate from the disease as well as the rate at which infected individuals can be found and isolated, does not change. In this case, a single equation for $\sig_1$ is required, such that (\ref{LogisticFunction}) is sufficient input to determine the latter as a function of time. Since in this case $\rhe$ is a constant free parameter, we consider $\rhe\in[0.04,0.1]$, which is roughly the range $\rhe(t)$ coverd in the right panel of Figure~\ref{Fig:TimeDepRates}. The result, in comparison to $\sig_1$ obtained assuming a time-dependent $\rhe$ are shown in Figure~\ref{Fig:TimeDepSigConstEps}.

\begin{figure}[htbp]
\begin{center}
\includegraphics[width=7.5cm]{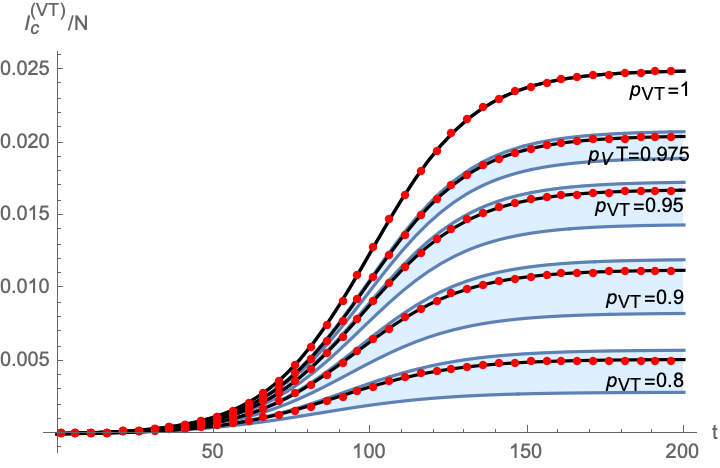}\hspace{1cm}\includegraphics[width=7.5cm]{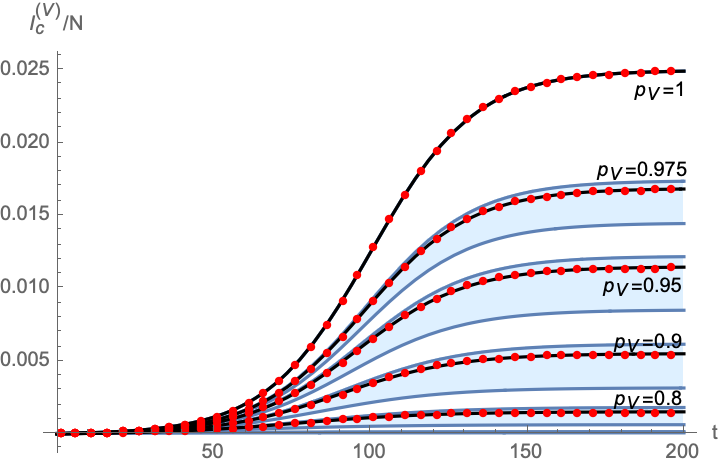}
\end{center}
\caption{$\Ic$ as a function of time for different values of $p$ using the \vtep\, model (\ref{DiffSIIRVgreenpass3G}) (left panel) or the \vep\, model (\ref{DiffSIIRVgreenpass}) (right panel). The blue bans correspond to the choice of $\rhe\in[0.04,0.1]$, while the red dots give the result of the computation assuming time-dependent $\rhe$ , along with their interpolation using a logistic function. Both plots use $A=0.025$, $\lambda=0.06$, $t_0=100$, $\delta=14$, $\rvac=0.0003$, $\riv=0.1$, $\sig_2/\sig_1=1$ and $\mv_0=0.5$. The interpolating black lines correspond to approximations with logistic functions.}
\label{Fig:TimeDepSigConstEps}
\end{figure}

Using this time-dependence to implement the \ep models (\ref{DiffSIIRVgreenpass3G}) and (\ref{DiffSIIRVgreenpass}) leads to different cumulative numbers of infected individuals as a function of time for different values of $\pv$ and $\pvt$, as is shown in Figure~\ref{Fig:TimeDepSigConstEps}, once more in comparison to the result of the main text which assumed a time-dependent $\rhe$. The blue bands in this plot represent the range obtained from $\rhe\in[0.04,0.1]$. Notice, that by design all choices of $\rhe$ reproduce the same function for $p=1$. Furthermore, for all choices of $\rhe$ and $p$, $\Ic/N$ can be approximated by a logistic functions. We note that these plots reveal a qualitatively similar picture to the computations in the main part of the paper.

\section{\ep\,in Different European Countries}\label{App:ExamplesPass}
In this appendix we provide a short overview over different \ep\, that have been discussed or introduced in some European countries. Our focus is on Austria, Denmark, France and Italy, which we have studied as examples in the main body of this paper.

\begin{itemize}
\item {\bf EU Digital COVID Certificate:} The EUDCC (also called Green Pass) was introduced in the European Union on 01/07 with the goal to facilitate travel within its 27 member states as well as Switzerland, Iceland, Norway, and Liechtenstein \cite{EUCOV}. It is a \vtep\, and is issued to individuals, who have been vaccinated, have recently tested negative for \cov\,(negative PCR test) or have recently recovered from a COVID-19 infection. It exempts individuals from further testing or quarantine when traveling within the participating countries.
\item {\bf Austria}: Austria has introduced a \vtep\, model as early as 19/05, whose range of applicability has has since been changed several times: following the '3-G-Regel' (geimpft, getestet, genesen) \cite{AustMinisterium} unrestricted access to public life is granted to individuals only who are either vaccinated, have recently tested negative for \cov\,(PCR test not older than 72 hours or antigen test not older than 48 hours) or have recovered from a previous COVID-19 infection (medical attestation not older than 6 months or a test for antibodies not older than 3 months). Restrictions in particular apply to establishments of the tourism industry, gastronomy and the attendance of public events. A plan unveiled by the government on 08/09 provides further restrictions to take effect as a function of the number of hospitalised individuals. 
\item {\bf Denmark:} In Denmark, a coronapas (or corona-passport), a \vtep\, has been introduced in the end of April \cite{coronapas}, which is required for Health to nightclubs and discotheques, as well as public events. Its applicability has changed since then and is currently for example no longer required for visiting restaurants and caf\'es. Individuals eligible for a corona-passport need to either prove that they have been vaccinated, have previously recovered from COVID-19, or have recently tested negative for \cov\, (PCR test not older than 96 hours or rapid antigen test not older than 72 hours).

\item {\bf France:} In France a 'pass sanitaire' was introduced on 01/06, first for events with more than 1000 participants, which was later gradually extended to cover mover and more aspects of public life \cite{FranceCOV}. Currently, individuals can obtain a 'pass sanitaire' if they have been completely vaccinated, have tested negative for \cov\,(PCR test not older than 72 hours or a supervised self-test), or have recently recovered from COVID-19 (positive test not older than 6 months).
\item {\bf Italy:} In Italy, since 06/08, Health to public places and cultural events (stadiums, museums, theaters, cinemas, \emph{etc.}) as well as sport centers (swimming pools and gyms) requires a 'green pass' \cite{ItalyCOV}, which is an extension of the EU's digital COVID certificate. Currently, the pass is not required for long distance travel across the country.

\end{itemize}

\bibliographystyle{ieeetr}
\bibliography{biblio}

\begin{thebibliography}{10}

\bibitem{cacciapaglia2021epidemiological}
G.~Cacciapaglia, C.~Cot, A.~de~Hoffer, S.~Hohenegger, F.~Sannino, and
  S.~Vatani, ``Epidemiological theory of virus variants,'' 2021.

\bibitem{MLvariants}
A.~de~Hoffer, S.~Vatani, C.~Cot, G.~Cacciapaglia, F.~Conventi, A.~Giannini,
  S.~Hohenegger, and F.~Sannino, ``Variant-driven multi-wave pattern of
  covid-19 via machine learning clustering of spike protein mutations,'' {\em
  medRxiv}, 2021.

\bibitem{ABC}
G.~Cacciapaglia, C.~Cot, M.~Della~Morte, S.~Hohenegger, F.~Sannino, and
  S.~Vatani, ``{The field theoretical ABC of epidemic dynamics},'' 1 2021.

\bibitem{Grassberger1983}
P.~Grassberger, ``On the critical behavior of the general epidemic process and
  dynamical percolation,'' {\em Mathematical Biosciences}, vol.~63, no.~2,
  pp.~157 -- 172, 1983.

\bibitem{Cardy_1985}
J.~L. Cardy and P.~Grassberger, ``Epidemic models and percolation,'' {\em
  Journal of Physics A: Mathematical and General}, vol.~18, pp.~L267--L271, apr
  1985.

\bibitem{Essam}
J.~W. Essam, ``Percolation theory,'' {\em Rep. Prog. Phys.}, vol.~43, p.~833,
  1980.

\bibitem{Pruessner}
G.~Pruessner, ``Field theory notes, chapter 6,'' {\em
  wwwf.imperial.ac.uk/$\sim$pruess/publications/Gunnar\_Pruessner\_field\_theory\_notes.pdf}.

\bibitem{Kermack:1927}
W.~O. Kermack, A.~McKendrick, and G.~T. Walker, ``{A contribution to the
  mathematical theory of epidemics},'' {\em Proceedings of the Royal Society
  A}, vol.~115, pp.~700--721, 1927.

\bibitem{HETHCOTErev}
H.~W. Hethcote, ``The mathematics of infectious diseases,'' {\em SIAM Review},
  vol.~42, no.~4, 2000.

\bibitem{DellaMorte:2020wlc}
M.~Della~Morte, D.~Orlando, and F.~Sannino, ``{Renormalization Group Approach
  to Pandemics: The COVID-19 Case},'' {\em Front. in Phys.}, vol.~8, p.~144,
  2020.

\bibitem{DellaMorte:2020qry}
M.~Della~Morte and F.~Sannino, ``{Renormalisation Group approach to pandemics
  as a time-dependent SIR model},'' {\em Front. in Phys.}, vol.~8, p.~583,
  2021.

\bibitem{Wilson1}
K.~G. Wilson, ``{Renormalization group and critical phenomena. 1.
  Renormalization group and the Kadanoff scaling picture},'' {\em Phys. Rev.
  B}, vol.~4, pp.~3174--3183, 1971.

\bibitem{Wilson2}
K.~G. Wilson, ``{Renormalization group and critical phenomena. 2. Phase space
  cell analysis of critical behavior},'' {\em Phys. Rev. B}, vol.~4,
  pp.~3184--3205, 1971.

\bibitem{Banks}
T.~Banks and A.~Zaks, ``{On the Phase Structure of Vector-Like Gauge Theories
  with Massless Fermions},'' {\em Nucl. Phys. B}, vol.~196, pp.~189--204, 1982.

\bibitem{cacciapaglia2020evidence}
G.~Cacciapaglia and F.~Sannino, ``Evidence for complex fixed points in pandemic
  data,'' {\em Front. Appl. Math. Stat.}, vol.~7, p.~659580, 2021.

\bibitem{cacciapaglia2020second}
G.~Cacciapaglia, C.~Cot, and F.~Sannino, ``Second wave covid-19 pandemics in
  europe: A temporal playbook,'' {\em Sci Rep}, vol.~10, p.~15514, 2020.

\bibitem{cacciapaglia2020us}
G.~Cacciapaglia, C.~Cot, A.~S. Islind, M.~{\'O}skarsd{\'o}ttir, and F.~Sannino,
  ``Impact of us vaccination strategy on covid-19 wave dynamics,'' {\em
  Scientific Reports}, vol.~11(1), pp.~1--11, 2021.

\bibitem{VaccDelta1}
J.~Lopez~Bernal, N.~Andrews, C.~Gower, E.~Gallagher, R.~Simmons, S.~Thelwall,
  J.~Stowe, E.~Tessier, N.~Groves, G.~Dabrera, R.~Myers, C.~N. Campbell,
  G.~Amirthalingam, M.~Edmunds, M.~Zambon, K.~E. Brown, S.~Hopkins, M.~Chand,
  and M.~Ramsay, ``Effectiveness of covid-19 vaccines against the b.1.617.2
  (delta) variant,'' {\em New England Journal of Medicine}, vol.~385, no.~7,
  pp.~585--594, 2021.

\bibitem{Cacciapaglia:2021vvu}
G.~Cacciapaglia, C.~Cot, M.~Della~Morte, S.~Hohenegger, F.~Sannino, and
  S.~Vatani, ``{The field theoretical ABC of epidemic dynamics},'' 1 2021.

\bibitem{Cacciapaglia:2020mjf}
G.~Cacciapaglia and F.~Sannino, ``{Interplay of social distancing and border
  restrictions for pandemics (COVID-19) via the epidemic Renormalisation Group
  framework},'' {\em Sci Rep}, vol.~10, p.~15828, 5 2020.

\bibitem{Worldometer}
Worldometer, ``Coronavirus cases.'' https://www.worldometers.info/coronavirus/,
  2021.

\bibitem{RKI}
R.~Koch-Institut, ``Robert koch-institut.'' https://www.rki.de/EN/Home, 2021.

\bibitem{AustMinisterium}
P.~u.~K. Bundesministerium~Soziales, Gesundheit, ``Coronavirus - aktuelle
  massnahmen.''
  https://www.sozialministerium.at/Informationen-zum-Coronavirus/, 2021.

\bibitem{HIT}
G.~P. Garnett, ``{Role of Herd Immunity in Determining the Effect of Vaccines
  against Sexually Transmitted Disease},'' {\em The Journal of Infectious
  Diseases}, vol.~191, pp.~S97--S106, 02 2005.

\bibitem{EUCOV}
BBC, ``The eu vaccine 'passport' and what it means for travel.''
  https://www.bbc.com/news/explainers-57665765, 2021.

\bibitem{coronapas}
BBC, ``Coronapas: The passport helping denmark open up after covid.''
  https://www.bbc.com/news/world-europe-56812293, 2021.

\bibitem{FranceCOV}
French\phantom{x}Gouvernment, ``Informations coronavirus.''
  https://www.gouvernement.fr/info-coronavirus, 2021.

\bibitem{ItalyCOV}
T.~Guardian, ``Italy imposes ‘green pass’ restrictions on unvaccinated
  people.''
  https://www.theguardian.com/world/2021/jul/22/italy-covid-19-green-pass-vaccinations-restrictions,
  2021.

\end{thebibliography}

\end{document}